\documentclass[preprint,showpacs,preprintnumbers,amsmath,amssymb,pra,endfloats*]{revtex4-1}

\usepackage{graphicx}
\usepackage{dcolumn}
\usepackage{bm}
\usepackage[bookmarks]{hyperref}

\begin{document}

\preprint{PRA}

\title{Determination of the effective transverse coherence of the neutron
wave packet as employed in reflectivity investigations of
condensed matter structures \\Part I: Measurements}

\author{Charles F. Majkrzak}
  \email{charles.majkrzak@nist.gov}
\author{Christopher Metting}%
\author{Brian B. Maranville}
\author{Joseph A. Dura}
\author{Sushil Satija}
\author{Terrence Udovic}
\author{Norman F. Berk}
\affiliation{%
Center for Neutron Research, National Institute of Standards and Technology\\
Gaithersburg, MD 20899, USA
}%

\date{\today}

\begin{abstract}
The primary purpose of this investigation is to determine the effective coherent
extent of the neutron wave packet transverse to its mean propagation vector $\mathbf k$,
when it is prepared in a typical instrument used to study the structure of 
materials in thin film form via specular reflection.  There are two principal 
reasons for doing so.  One has to do with the fundamental physical interest in 
the characteristics of a free neutron as a quantum object while the other is of 
a more practical nature, relating to the understanding of how to interpret 
elastic scattering data when the neutron is employed as a probe of condensed 
matter structure on an atomic or nanometer scale.  Knowing such a basic physical 
characteristic as the neutron's effective transverse coherence can dictate how 
to properly analyze specular reflectivity data obtained for material film 
structures possessing some amount of in-plane inhomogeneity.  In this study we 
describe a means of measuring the effective transverse coherence length of the 
neutron wave packet by specular reflection from a series of diffraction gratings 
of different spacings.  Complementary non-specular measurements of the widths of 
grating reflections were also performed which corroborate the specular results.  
(Part I principally describes measurements interpreted according to the 
theoretical picture presented in Part II.)  Each grating was fabricated by 
lift-off photo-lithography patterning of a nickel film (approximately 1000 
Angstroms thick) formed by physical vapor deposition on a flat silicon crystal 
surface.  The grating periods ranged from 10 microns (5 microns Ni stripe, 
5 microns intervening space) to several hundred microns.  The transverse 
coherence length, modeled as the width of the wave packet, was determined from 
an analysis of the specular reflectivity curves of the set of gratings.
\end{abstract}

\maketitle

\section{\label{sec:introduction}Introduction}
\subsection{\label{subsec:plane_waves}Plane waves}
Historically, the theoretical framework for many neutron elastic scattering studies of the structure of condensed matter has been based on a model of the scattering of incident beams of plane waves, each neutron in the beam described by a plane wave state of definite energy and momentum, but allowing for a distribution of energy and momentum among the distinguishable incident neutrons.  This distribution is then accounted for in the data analysis in terms of an instrumental resolution function, which incorporates the uncertainty of the state of each individual neutron in the beam.

If the scattering is sufficiently weak (e.g., as in diffraction studies of polycrystalline materials or in small angle scattering from dilute solutions), then the Born approximation, in which the scattering length density (SLD) distribution of the material is related to the reflected neutron wave function by a Fourier transform, can be accurately applied.  On the other hand, if the scattering is strong enough, solution of the time-independent, one-dimensional Schroedinger equation is required (e.g., for specular neutron reflectometry).  However, in both of these cases, the incident and reflected wave functions are still normally taken to be plane waves.  Thus, the coherence of each and every individual neutron in the beam is implicitly assumed to be infinite and the broadening of, say, an observed Bragg reflection beyond its natural correlation width (determined by the number of equally spaced atomic planes contributing in perfect periodic order) is attributed to the width of the instrumental resolution function along the corresponding direction of wavevector transfer.

\subsection{\label{subsec:importance}Importance of coherence in specular neutron reflectometry}
In practice, it turns out that the process of elastic, specular neutron reflection (SNR) (wave vector and momentum transfer perpendicular to the surface of the film) from layered film structures -- which have in-plane density variations on short enough length scales and/or of sufficiently small magnitude  -- is remarkably well-described (i.e., with a quantitative accuracy of the order of one percent for cases approaching ideal conditions) by plane wave solutions of a time-independent, one-dimensional Schroedinger wave equation.  Nonetheless, it is unphysical, in principle, to describe the neutron wave function as a single plane wave which is composed of identical, parallel wave fronts of infinite lateral extent.  In reality, the wavefronts representing an incident neutron are of finite size (i.e., perpendicular to its wavevector).  Consequently, the adoption of a single plane wave as the solution of the wave equation describing SNR is valid in a theoretical analysis only if the in-plane sample area illuminated by the projection of the transverse extent of the wave onto the surface is sufficient to effectively average over any existing in-plane inhomogeneities in density.  More specifically, for a valid SLD depth profile along the surface normal to be extracted from SNR data, the spatial extent of the in-plane variations in SLD must be small enough to ensure that each neutron effectively averages over those fluctuations.  Otherwise, an area-weighted, incoherent sum of independent reflectivities would be measured, along with partially-averaged components, which must be dealt with accordingly.  If such incoherence were not recognized, analyzing the data as though it were associated with a single SLD profile would be incorrect.

It was discovered previously \cite{majkrzak2000first}, that even when the coherence of the neutron is not known explicitly, in-plane inhomogeneities larger than its effective transverse coherence length can be manifest, for specular reflection, in the imaginary part of the reflected wave function -- which can be determined by phase-sensitive methods employing reference layers \cite{majkrzak2003phase}.  Nonetheless, the evidence obtained in this way is more of diagnostic value and does not directly provide quantitative information about the relevant spatial length scales of either the probe or scattering object of interest.

\subsection{\label{subsec:functions}Coherence functions}
Within the last decade or so, efforts have been made to apply the theoretical treatment of partially coherent beams in light optics \cite{born1975principles,mandel1995optical,van1934wahrscheinliche,zernike1938concept} to neutron beams \cite{felber1998coherence,gahler1998space,de2008coherence,de2010coherence} through the use of appropriate correlation functions, taking into account the effects of instrumental components such as slits, monochromating crystals, etc., which ultimately define the relevant coherence volume.  The treatment of coherence in neutron interferometry has been especially important given the nature of the fundamental concepts in quantum mechanics which are investigated by this means -- consideration of this particular aspect has been addressed, for example, by Rauch and Werner \cite{rauch2000neutron} where they include a discussion of an approach to describe the coherence phenomena of quantum fields based on an auto-correlation function introduced by Glauber \cite{glauber1963quantum,glauber1963coherent}.  An appreciable amount of work also has been done to extend the concept of partial coherence originally developed for visible light to x-ray photons -- in particular to the different types of x-ray beams produced by various devices at synchrotron sources (see, for example, \cite{sinha1998effects,bernhoeft1998probe,gutt2008effects}).

\subsection{\label{subsec:source_significance}Significance of the nature of the source}
Exactly how coherence is treated mathematically -- for example, in terms of a correlation function as mentioned above -- must depend on the relevant properties of the source involved.  For example, a continuously oscillating macroscopic electromagnetic field produces radiation that is fundamentally different in many regards than that emitted by a laser where the photons emerge in a highly-correlated state.  And a source which produces neutrons individually by nuclear fission reaction and subsequent moderation by liquid hydrogen, for instance, is different from either of the aforementioned electromagnetic sources.  Due consideration is required to construct a mathematical function which unambiguously describes a particular property of the radiation emitted, e.g., its coherence, distinctly from those characteristics intrinsic to the source.

To illustrate this point, consider the minimalist schematic of Figure 1 which represents the effect of illuminating a pair of apertures by two independent, incoherent (spatially and temporally), and sufficiently weak sources, one of which is displaced off the central axis of symmetry from the other.  
\begin{figure}
\includegraphics[width=\linewidth]{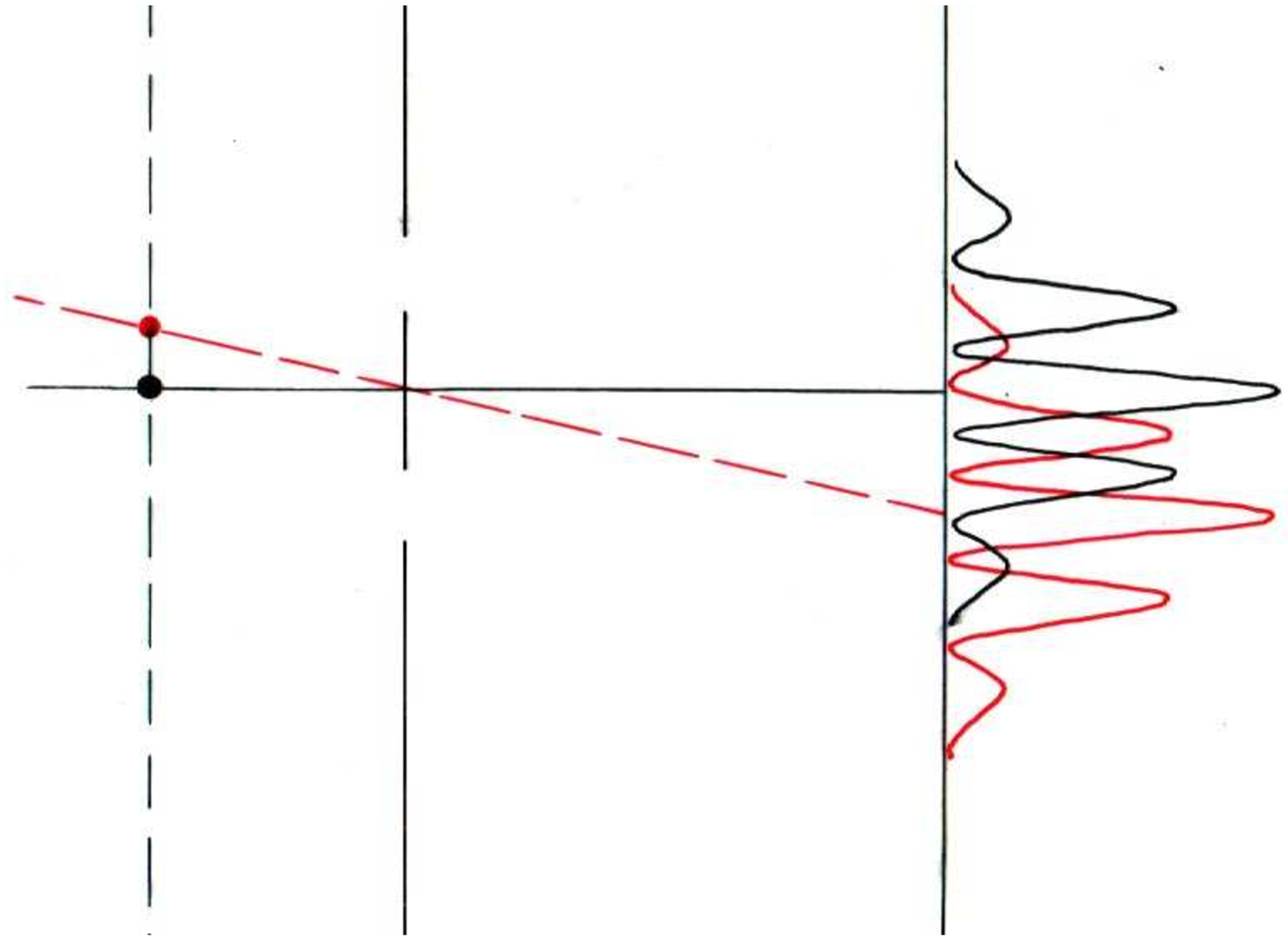} 
\caption{\label{fig:1} 
 (Color online)  A schematic depicting: on the left -- a pair of spatially and temporally incoherent source points (one of which is displaced off the axis of symmetry); in the middle -- a pair of apertures in an otherwise opaque barrier; and, on the right -- a detecting screen.  From either source emanate individual localized wave forms or packets, each having sufficiently small volumes so as not to overlap or interact with one another, yet large enough and of a composition that ensures a completely coherent interaction with the pair of apertures.  The separate diffraction patterns associated with each of the two source points -- which {\it{would have been}} observed on the detecting screen -- are shown -- had their corresponding intensity distributions {\it{not}} been superimposed.  The composite image which {\it{would}} actually be seen represents a sum of the intensities of the two independent, underlying diffraction patterns.  In this particular case, it is the spatial separation of the source points which effectively obscures, to a degree that depends upon the distance between source points, the original single-source-point diffraction pattern features -- in no way does the resultant pattern indicate a loss of coherence within any individual localized wave packet.  Conversely, any reduction in ``fringe visibility'' is due entirely to the incoherent effect associated with the extension of the source.}
\end{figure}
Imagine that each source point emanates one quanta of radiation at a time, localized enough in a finite volume of space and time so as not to overlap or interact with any other quanta -- yet of sufficient lateral and longitudinal (perpendicular and parallel to the wave propagation vector, respectively) spatial dimensions that each quanta interacts with the diffracting object (the {\it{pair}} of slits) in an almost perfectly coherent manner.  (These localized wave forms are traditionally referred to as wave packets and will be discussed further in subsequent sections.)  After a statistically significant number of individual quanta from each (of the two) sources have scattered from the double-slit, a pair of distinct, perfectly resolvable, interference patterns are created -- however, the separate {\it{intensity}} distributions for these two patterns  are superimposed on one another on the detecting screen downstream, thereby producing an observed net result in which the features, such as interference fringe maxima and minima, associated with either of the original underlying patterns are less well-defined.  This so-called loss of ``fringe visibility'' (see, for example, Section 12.2 in \cite{hechtoptics1998}) in this circumstance, however, has nothing to do with loss of coherence as embodied in any one of the distinct, independent radiation quanta which contribute, collectively, to the creation of the diffraction pattern associated with one or the other source point.  To the contrary, the net pattern observed on the detecting screen is indicative only of the completely incoherent effect corresponding to the extension of the source (i.e., as represented in this simple case by the two separate point sources).

\subsection{\label{subsec:model_source}Model for a neutron source}
It is widely believed that neutrons emerge from a hydrogen moderator as individual, mutually incoherent spherical waves, in essence the waves produced by the last (largely spin-incoherent, isotropic) interaction that each neutron undergoes.  If the spherical wave associated with each neutron were to propagate undisturbed, its wavefronts would become more planar as distance from the source increased.  However, along the way the wave can become further shaped (and localized) by entry into a guide tube, subsequent reflection from guide surfaces and a monochromating crystal, transmission through apertures, and even reflection from the substrate on which a sample film to be studied is deposited.  Imagine the distortion to a wave front which might occur upon reflection from a guide surface that resembled a wavy circus mirror on some appropriately small length scale -- or, as the neutron undergoes Bragg diffraction from a micron-size mosaic block in a  monochromating crystal, the further modification to both the longitudinal and transverse extent of the localized wave packet being formed in space.  So too, if the neutron passes through a pair of rectangular apertures, the wave packet shape and composition will be altered, to a degree depending on the spatial dimensions of the aperture relative to the neutron wave packet's mean wavelength (which can be described, as appropriate, by either Fraunhofer or Fresnel diffraction theory \cite{sears1989neutron}).  However, these very same optical elements in the experimental set-up, i.e., moderator, guides, monochromating crystal, and slits also define, at the same time, the incoherent distribution of mean wavevector values corresponding to each of the individual neutrons which constitute the beam.  

\begin{figure}
\includegraphics[width=\linewidth]{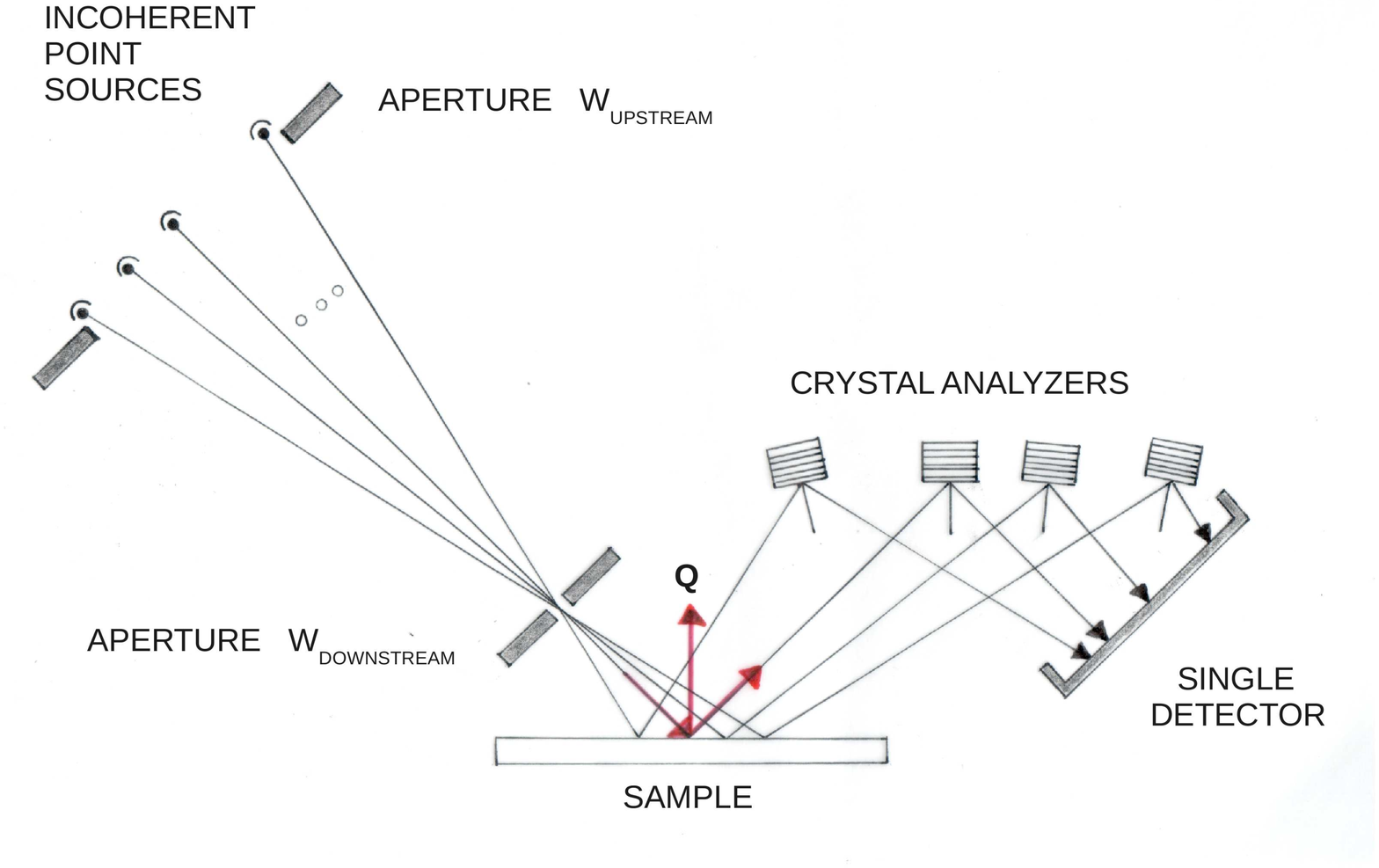} 
\caption{\label{fig:2} 
(Color online)  Essential components of a reflectometer that are pertinent for forming the coherence of an individual neutron wave packet as well as the distributions of the directions (angles) and magnitudes of the mean wavevectors of the collection of wave packets making up a beam.  The width of the ``upstream'' aperture adjacent to the source is significantly wider than that of the ``downstream'' slit near the sample.  The source points, an arbitrary number of which are depicted, are taken to be individual hydrogen atoms, the nucleus of each serving as a spin incoherent scattering source of an isotropic outgoing spherical wave representing a single neutron -- each hydrogen atom scatters an incident neutron independently, both spatially and temporally, of all the other hydrogen atoms composing an extended source.   After subsequent elastic, specular reflection from a flat sample, each neutron is Bragg reflected from a perfect microcrystallite of an ideally imperfect crystal such as pyrolytic graphite.  No matter what trajectory a given reflected neutron takes through the instrument, it is captured in the same detector without registering its location of entry and therefore encodes no knowledge of angle other than that defined by the overall width of the detector and aperture preceding the sample.  See text for discussion of how, in certain circumstances, it is possible to, at least partially, separate the uncertainty associated with the intrinsic coherent superposition of constituent wavevectors composing an individual neutron wave packet from that of the incoherent distribution of the mean wavevectors of the packets forming a beam.}
\end{figure}

As an example, consider Figure \ref{fig:2} which depicts the essential components of a reflectometer that are pertinent for forming the coherence of an individual neutron wave packet as well as the distributions of the directions (angles) and magnitudes of the mean wavevectors of the collection of wave packets making up a beam.  In this particular configuration, the monochromating crystals are located after the neutrons are specularly reflected from the sample -- although this is not necessarily the most common arrangement, it is useful for illustrative purposes.

Assume that the width of the ``upstream'' aperture adjacent to the source is significantly wider than that of the ``downstream'' slit near the sample.  The source points, an arbitrary number of which are depicted, are taken to be individual hydrogen atoms, the nucleus of each serving as a spin incoherent scattering source of an isotropic outgoing spherical wave representing a single neutron -- each hydrogen atom scatters an incident neutron independently, both spatially and temporally, of all the other hydrogen atoms composing an extended source.  (In the figure, the semicircular boundary drawn around each source point, the size of a nucleus, is to emphasize the independence of that source.  As in the case for absorption of a neutron by a nucleus, in the spin-incoherent scattering process a neutron interacts with a single nucleus regardless of whether its incident wavefunction spans a region containing other such nuclei.)  After subsequent elastic, specular reflection from a flat sample, each neutron is Bragg reflected from a perfect microcrystallite of an ideally imperfect crystal such as pyrolytic graphite (along the (002) crystallographic direction).  Such crystallites are typically of dimensions of the order of 1 to 10 microns and coherently reflect over a corresponding volume (because the number of atomic planes contributing to the reflection is relatively large, the width of the Bragg reflection at a single specified wavelength is, consequently, relatively narrow, typically a second of arc or less).  This type of mosaic crystal has an angular distribution of mosaic block normals of the order of half a degree FWHM.  No matter what trajectory a given reflected neutron takes through the instrument, it is captured in the same detector without registering its location of entry and therefore encodes no knowledge of angle other than that defined by the overall width of the detector and aperture preceding the sample.

As we have discussed previously, one neutron traverses the reflectometer at a time, independent of all the other neutrons composing a beam.  The direction of the mean wavevector of the packet representing a given neutron is determined (for the present configuration) by the location of the source point and the width of the downstream aperture -- the upstream aperture is so wide in comparison that it has negligible affect in shaping a packet.  The magnitude of its mean wavevector, on the other hand, is determined by its angle of incidence with respect to the mosaic block of the analyser crystal.  The uncertainty principle establishes a limiting relationship between the spatial transverse extent of the wave packet and its intrinsic momentum distibution.  In terms of wavevector, this can be stated for the transverse component $k_\perp$ as $\Delta k_\perp \Delta r_\perp \geq 1/2$ or $\Delta r_\perp \geq 1/(2\Delta k_\perp)$. 

Now $\Delta k_\perp$  is the uncertainty in the packet's wavevector transverse to its mean propagation direction and can be computed in principle from the angular divergence defined by the point source, downstream slit, analyser mosaic crystallite block size and orientation (again, presuming elastic specular reflection from a perfectly flat sample).  (Incidentally, the angular divergence defined by a point source and the 
downstream slit is given by the arc tangent of one half the slit width divided by the distance between source and slit -- only a factor of two smaller than two slits of the same width the same distance apart.)  From the Figure it should be clear that no matter which source point across the extended opening defined by the upstream aperture a given neutron emanates from, its $\Delta k_\perp$  and, thus, $\Delta r_\perp$ , will be nearly the same.  Note, however, that this $\Delta k_\perp$ is demonstrably \textit{not} necessarily the wavevector uncertainty associated with the distribution of mean wavevector values of the packets composing the beam -- in other words, the mean wavevector transfer $Q$ associated with a given single neutron described as a wave packet can be appreciably different depending upon the particular position across the width of the upstream aperture that the neutron originates from.  To express it in other words, one can imagine an effective miniature aperture at each individual source point which, in conjunction with the common downstream slit near the sample, defines a subset of trajectories that are but one contribution to the overall beam divergence resulting from the incoherent sum of contributions from all source points -- whereas each individual subset defines the wave packet coherence of a single neutron.

This \textit{incoherent} instrumental \textit{beam} resolution $\Delta Q$ is described in terms of the angular divergence allowed by the average of the upstream and downstream slit widths in conjunction with the properties of the analyser crystal.  Because of the relatively low grazing angles of incidence typical of specular reflectometry measurements, the uncertainty in the transverse component of the distribution of mean wavevector values associated with the ensemble of packets composing the beam is approximately given by $\Delta Q$ where the wavevector transfer $Q \approx 2 k_\perp$.  The appropriate expression for $\Delta Q$ can be written in terms of the beam divergence and wavelength bandwidth, as given in Equation \ref{eq:12} in Section \ref{sec:specular_analysis} below.  It is, therefore, \textit{not} in general correct to equate $\Delta r_\perp$ with $1/ \Delta Q$ -- as the measurements described in subsequent sections confirm.

As shown in the sections which follow, the analysis and interpretation of the observed reflection (both specular and nonspecular) of a neutron from a grating structure is consistent with there being a certain amount of separation between the coherence properties corresponding to an individual neutron wave packet and the macroscopic characteristics representing a collection of similar neutrons which compose a beam -- namely, the angular divergence and wavelength spread associated with the distribution of the mean wavevectors of the wave packets in the ensemble.  This should not be surprising given the nature of the source -- a liquid hydrogen moderator wherein each interaction between hydrogen nucleus and neutron is incoherent and isotropic and emission occurs randomly at different times and locations.  Moreover, in a well-collimated, quasi-monochromatic incident beam on a typical neutron reflectometer, a flux of $10^5 \frac{\mathrm{n}}{\mathrm{cm}^2 \cdot \mathrm{s}}$ implies an average spacing of about 1 cm between successive neutrons along the nominal beam direction (at a wavelength of 4 Angstroms) -- a macroscopic distance which, given the wave packet dimensions inferred from the results reported in the present work, would not be consistent with appreciable overlap of the wave functions of any two independent neutrons.

And so, one physical picture of the type of neutron source which we are considering here may be described as follows.  A collection of neutrons emerges from a spatially and temporally incoherent hydrogen moderator source and is further formed by guides, monochromator, and apertures into an incident beam composed of independent, non-interacting neutron wave packets.  Each neutron wave packet then scatters from a material object on its own, giving rise to a characteristic reflectivity curve or diffraction pattern -- once a sufficient number of neutron wave packets in the beam have scattered from a material object to yield data of the requisite statistical accuracy.  It would be possible, in principle, to turn a nuclear fission reactor and liquid hydrogen moderator on, have a single neutron scatter from a diffracting object, shut the source down, exchange the uranium in the core and the liquid hydrogen in the moderator vessel, start up the source again and scatter another single neutron, repeating this sequence \textit{ad infinitum}, one neutron at a time, and eventually obtain the characteristic diffraction pattern unique to that object.  The remark attributed to Dirac \cite{dirac1958principles} that ``. . . a photon only interferes with itself . . .'' may not apply in every circumstance (e.g., for the highly-correlated, many-body entangled state of light emitted by a laser) but for the neutrons being considered here -- we believe it does.  (If any doubt were to arise regarding the possibility of overlap of individual neutron wave functions, it would be straightforward to perform diffraction or reflection experiments with differing beam intensities [through the simple use of a set of purely absorbing attenuators of varying degree] to detect such an effect.)  The present case is analogous to that of a sufficiently weak and completely incoherent light source of optical wavelength photons for which interaction with, say, a double slit of appropriate dimensions has been demonstrated to give rise to a characteristic diffraction pattern -- even at such low incident beam intensities that each photon appears on a detecting film one quantum at a time (examples such as this can be found in many introductory texts on optics e.g., \cite{hechtoptics1998} and quantum mechanics, e.g., \cite{anderson1971modern}).

Nonetheless, the neutron as wave packet view which we are adopting for the description of the present work is not without problems nor is it universally accepted.  Many of the issues of contention, such as the spread of some forms (e.g., Gaussian distribution of plane wave states) of a material wave packet with time, are considered in reference \cite{utsuro2010handbook} (which is also a reasonably comprehensive source for original research articles on the subject).  We believe, however, that the results of the experiments which we report here are useful and applicable  in regard to neutron reflectometry regardless of the interpretation of the physical meaning of a neutron wave packet or its detailed shape and composition.  The essential requirement is that an individual free neutron be described by some composite waveform which is a solution of the time-dependent Schroedinger equation and also localizes, to an appropriate extent, the volume of interaction of a single neutron with a material object in real space, independently of any other neutron in a beam.   

\subsection{\label{subsec:wavepacket}The neutron as a wave packet}

To more formally describe the wave-like behavior of a relatively localized quantum object, such as the neutron we have been considering, the concept of a \textit{wave packet} was conceived originally (see, e.g., \cite{cohenlaloe}) as a mathematical representation consisting of a continuous, coherent superposition of plane-wave momentum eigenstates, each associated with a specific wavevector $\mathbf k_j$ . (The distribution of the component plane wave states is frequently taken to be Gaussian because of its mathematical utility in theoretical considerations, but such a distribution is neither necessary nor the one most appropriate in a given circumstance.)  The wave packet construction effectively creates a probability distribution for finding a neutron in a finite, localized region of ordinary space while at the same time preserving the essential wave property of coherence.  Such a picture is also consistent with the Heisenberg uncertainty principle and, in fact, is a natural consequence of it \cite{cohenlaloe}.

Given, then, that each neutron wave packet is itself ultimately responsible for coherent diffraction processes, it is instructive to re-examine the underlying equation which describes the wave motion. The wave function ψ for an independent, single neutron in free space satisfies the Schroedinger equation

\begin{equation}
    \label{eq:1}
    \left\{[\hbar^2/(2m)] \nabla^2 + i\hbar (\partial/\partial t)\right\} \psi = 0
\end{equation} 
where a special or particular solution is the plane wave

\begin{equation}
    \label{eq:2}
    \psi(\mathbf r, t) = A \exp[i(\mathbf k \cdot \mathbf r - \omega_k t)]
\end{equation} 

An individual neutron in a single plane wave state would be totally coherent.  Although such a state is a mathematical idealization and is physically unattainable in practice, it is, as already mentioned above, still of great value and use in mathematical descriptions of specular reflectivity.  A more realistic and
general solution, on the other hand, can be expressed as a \textit{coherent} superposition of plane wave states or wave packet,

\begin{equation}
    \label{eq:3}
    \Psi(\mathbf r, t) = \int A(\mathbf k) \exp[i(\mathbf k \cdot \mathbf r - \omega_k t)] ~ d\mathbf k
\end{equation} 

Note that the amplitudes $A(\mathbf k)$ are the weighting factors which effectively determine the distribution that defines the size and shape of the wave packet representing the neutron as a quantum object -- a wave localized in space.  Whatever coherence properties are associated with an individual neutron are embodied in the coefficients or weighting factors of the plane wave momentum eigenstates of this distribution which constitute the wave packet.  The degree of coherence can have varied measures depending on the physical circumstances or instrumental configuration and can involve spatial and/or temporal -- transverse or longitudinal -- components (i.e., perpendicular or parallel to the mean value of the wave vector associated with the packet).  In elastic diffraction, it is the transverse extent of the wave packet as well as the shapes of and correlations between successive wavefronts contained within the packet that are of primary significance.  The shape and size of the reflecting material object in conjunction with the correlations existing within the packet determine the diffraction pattern that is observed (the surfaces of constant phase are not necessarily planar).  And so, the wave packet function $\Psi (\mathbf r,t)$ \textit{itself}, as given in Equation (\ref{eq:3}), describes the coherence of individual neutrons in an incoherent mixture that compose a beam (see, for example, \cite{rauch2000neutron}, p.95).  Note that the coherence properties contained within the wave packet are determined in a fundamental way through the requirement that $\Psi (\mathbf r,t)$ be a general solution of the \textit{time-dependent} wave equation (e.g., for a neutron with finite mass, the width of the wave packet spreads with time).  While the stationary state solutions of the Schroedinger equation and the stationary wave (single mode) solutions of Maxwell's equation for the transverse component of the electric field in a dielectric medium can be written in similar form (as is commonly done in unifying the description of neutron and x-ray elastic scattering), this mathematical equivalence does not extend to the general case of the corresponding time-dependent equations.

\subsection{\label{subsec:work_description}Description of the work reported on here}
Here we describe a means of determining one particular measure of the effective transverse coherence length of the neutron wave packet, as created within a typical instrument, that is relevant to the process of specular reflection.  The method involves the use of well-characterized diffraction gratings which serve as a probe of the neutron wave packet.  Although the source and instrumental components (such as the moderator, guides, monochromator, and apertures) determine the individual wave packet properties (size, shape, and coherence) as well as define, simultaneously, the collective beam characteristics, the technique we employ is largely independent of the incoherent distribution of mean packet wavevectors that constitute a beam.  In particular, it is shown that this measure of transverse coherence of a neutron wave packet is not necessarily inversely proportional to the wavevector uncertainty or incoherent instrumental resolution associated with the distributions of angular divergences and wavelengths in the beam.  Although the measure of transverse coherence which we extract is related to the longitudinal coherence of the packet (e.g., through the paraxial Helmholtz equation \cite{saleh1991fundamentals}),  we do not consider the latter here since the methods we employ are not directly sensitive to it (see, e.g., \cite{rauch2000neutron} for discussion of the longitudinal component). 

A theory of the scattering of wave packets is developed in the accompanying article, Part II, a brief overview of which is given in a separate section below.

\section{\label{sec:coherent_vs_incoherent}Distinguishing between wave packet coherence and incoherent beam distributions}
There are certain conditions and instrumental configurations for which it is possible to extract some measure of the coherence of a neutron wave packet as well as the distribution of the mean wavevector values $\mathbf k$ (angle and magnitude) associated with each wave packet composing a beam.  Consider, for example, the relatively simple arrangement for measuring elastic scattering from a material sample that is shown schematically in Figure 3a.

\begin{figure}
\includegraphics[width=\linewidth]{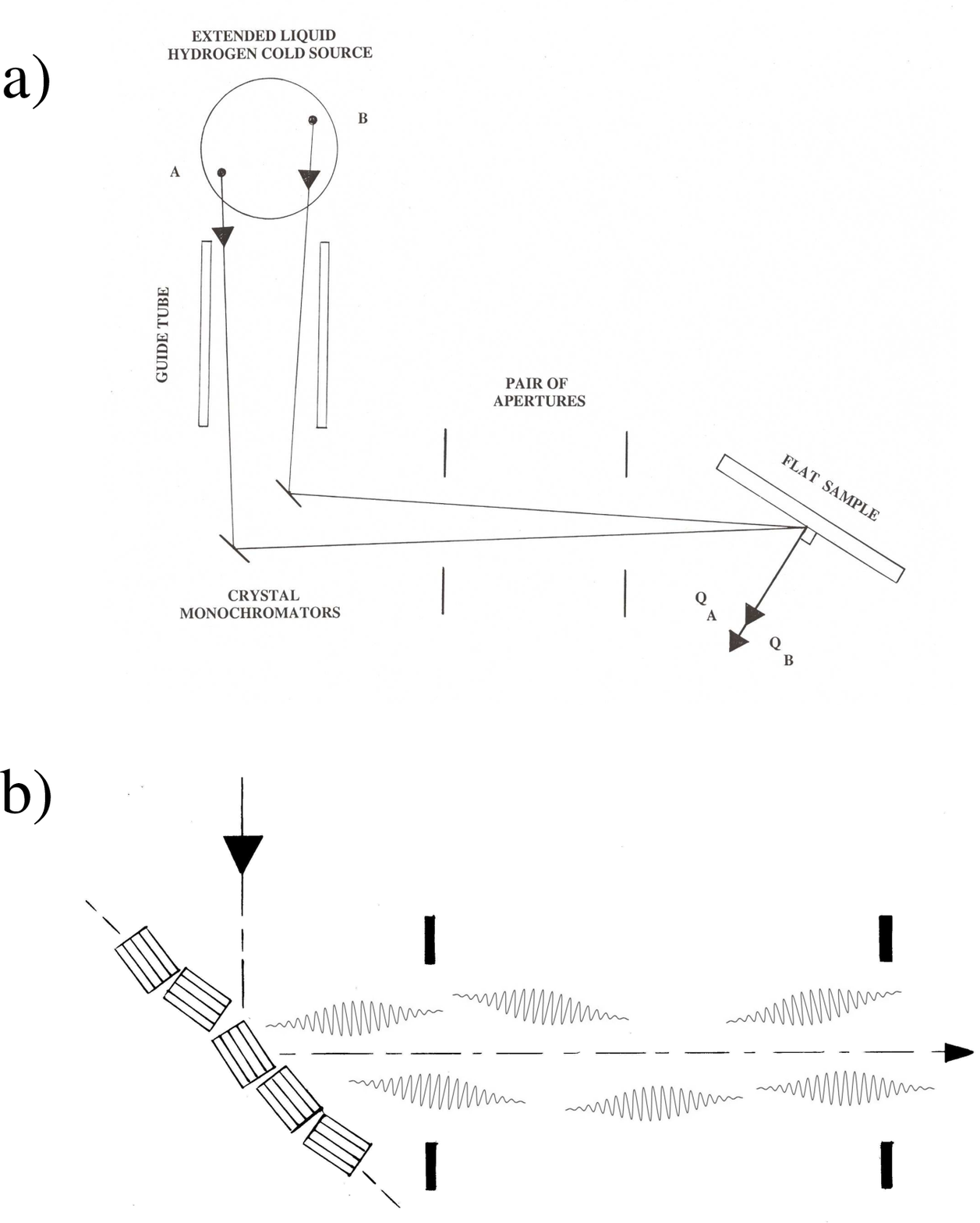}
\caption{\label{fig:3}
a) Simple instrumental arrangement for measuring elastic reflection from a material sample.  See related discussion in text. 
b) detail of a) depicting a mixture of wave packets composing a beam, along with the mosaic crystal monochromator structure which is presumed to contribute to determining the wave packet coherence.  The widths of the two apertures, on the other hand, are typically large enough that single-slit diffraction effects are negligible (as are effects from any single mask edge).  But both crystal and slits also help define the incoherent instrumental resolution of the beam of individual neutron wave packets.}
\end{figure}

Suppose neutron ``A'' in Figure \ref{fig:3}a has its last interaction with the liquid hydrogen moderator material by scattering isotropically and incoherently from a single hydrogen nucleus in the form of a spherical wave.  Assuming that it follows a path into the totally reflecting guide tube, it may undergo multiple specular mirror reflections from a Ni coated glass surface and be subsequently Bragg reflected from a single mosaic block of perfect crystal in a monochromator through a pair of apertures to be incident on some sample object to be studied.  In this set-up, the single crystal mosaic block can play a significant role in determining the size and shape of the coherent volume of the neutron wave packet and limiting the range of wavevectors corresponding to each of its plane wave components.  For a common monochromator material such as pyrolytic graphite, the size of such a mosaic block is of the order of 1 to 10 micrometers \cite{simonis2002stm}.  A typical slit width, however, for either of the pair of apertures, is of the order of 1 mm -- for such a width and a commonly used neutron wavelength of 5 Angstrom, diffraction effects are negligible and therefore should not appreciably contribute to shaping the neutron wave packet.  But at the same time, both the monochromator and slits also affect the distribution of mean values of the wavevectors of the mixture of wave packets making up a beam, as discussed earlier in the Introduction.

Another neutron, say ``B'', emanating from a different location on the extended moderator source but following an essentially equivalent trajectory to that of ``A'' can be expected to have a very similar wave packet shape, although with possibly a significantly different mean wavevector $\mathbf k$ within the range of geometrical angular divergence allowed for the beam by the pair of apertures between monochromator and sample.  The distribution of mean wavevector directions defined (primarily) by the slits is a measure of the incoherent instrumental angular resolution of the mixture of neutron packets within a beam.  Figure \ref{fig:3}b illustrates this picture schematically.

It will be shown, in the following sections, that by analyzing the shape of the specular reflection curve  for each of a set of uniform, rectangular gratings of incrementally differing periods -- particularly in the region about the characteristic critical angles below which total external neutron reflection occurs -- the effective transverse coherence length of the neutron wave packet can be determined within well-defined limits, almost independently of the angular divergence and wavelength spread of the beam.   This is possible because of the significant changes in the shapes of the specular reflectivity curve which result when the neutron wave packet effectively averages over the bars and troughs of a grating of a given period -- or not -- as proposed in reference \cite{majkrzak2006structural}.  A similar approach has recently been applied for x-rays \cite{lee2011determining}.  In addition, we have corroborated the results of the specular measurements by performing a more conventional analysis of the line widths of the grating reflections in off-specular scattering, given that the number of correlated periods of the grating (see section below describing the fabrication and characterization) extends a distance far greater than the projection of the transverse coherence length of the neutron or x-ray onto the reflecting surface of the sample object, as was done in reference \cite{salditt1994x}.

And so, in the experiments reported herein, the conventional roles of the neutron and grating are interchanged -- we use the gratings of ``known'' physical characteristics to probe an ``unknown'' wave property of the neutron.  Measurements were performed for various neutron monochromators and sources (pyrolytic graphite, perfect single crystal silicon, vanadium), different beam angular divergences and wavelength bandwidths, and, for comparison, x-rays produced by a fixed anode source.

\section{\label{sec:instrumental}Instrumental configuration and characteristics}
\begin{figure}
\includegraphics[width=\linewidth]{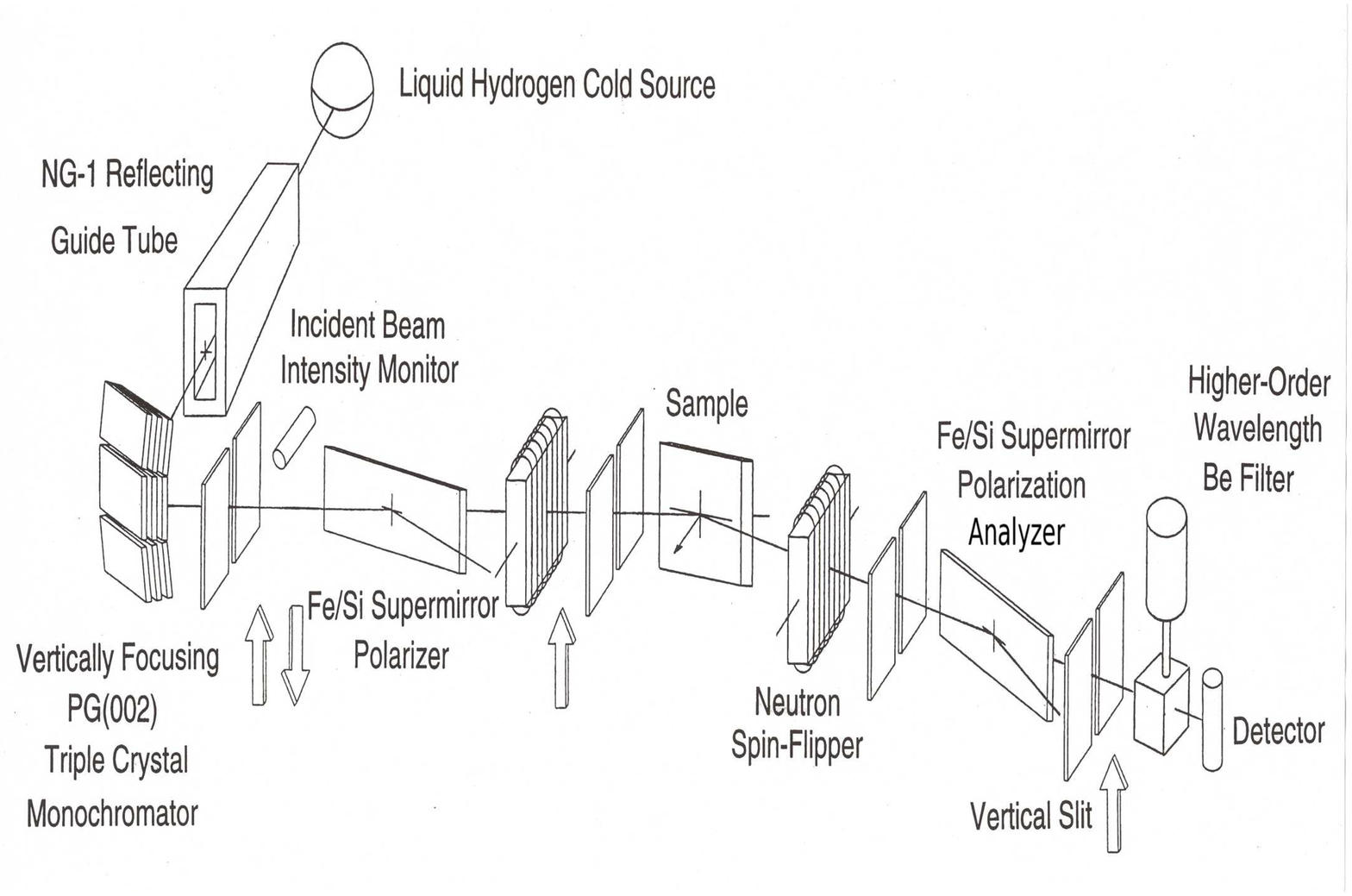}
\caption{\label{fig:4} 
Schematic of a polarized neutron reflectometer as described in the text.  For most of the reflectivity measurements performed  in the present work, the Be filter was located in between guide tube sections (not as indicated next to the detector), and an additional PG(002) analyzer crystal was located between the last slit and the detector in the ``W'' configuration for some measurements.  Other configurations, some not involving polarized beams, were also employed and will be described where relevant.  (Adaptation of Fig. 12.7 of \cite{majkrzak2006structural}.)}
\end{figure}

Figure \ref{fig:4} is a schematic of the reflectometer configuration employed in the specular and nonspecular reflectivity measurements performed on the grating structures (a similar instrument has been described previously in greater detail \cite{dura2006and}).  The neutrons were originally polarized so as to be more sensitive to the magnetization of the saturated Ni stripes used to construct the periodic rectangular gratings.  However, aspects of the measurements relating to the interaction of polarized neutron wave packets with the magnetization will not be dealt with in the present paper.

As discussed in earlier sections, with each neutron in the beam we associate a wave packet possessing a nominal or mean wavevector $\mathbf k$ .  A distribution of these nominal wavevectors constitute a beam within the reflectometer (either incident or reflected) which is defined primarily by the macroscopic apertures and mosaic crystal monochromator (pyrolytic graphite) of the instrument.  (Throughout the paper, uncertainties in quantities are expressed as full width at half maximum values, FWHM, assuming a corresponding Gaussian distribution.)  The apertures and width of the monochromating crystal's angular distribution of perfect single crystalline mosaic blocks (approximately 0.5 degrees FWHM) define an {\textit{incoherent}} instrumental resolution (IIR) component associated with the beam.  (As already mentioned, diffraction effects from the edges of the masks which define the apertures are negligible in the present context -- the slit dimensions are typically of the order of a mm.)  The vertical heights of the slits are relatively large, of the order of 10 cm, so that the vertical angular divergence of the beam is approximately 2.5 degrees FWHM.  For the majority of the measurements performed, the beam was focused in the vertical plane by the curvature of the composite PG(002) monochromator to a height of approximately 3 cm at the sample position.  In the horizontal scattering plane defined by incident and final neutron wavevectors, the angular divergences of the incident and reflected beams are typically of the order of 0.01 degree.  The wavevector magnitude resolution, determined by the monochromator atomic plane spacing and angular divergences allowed by the mosaic distribution, guide tube critical angles for total external reflection, and the apertures turns out to be of the order of 0.01 for typical values of the slit widths used in this study.  However, as presented below, measurements with other values of angular divergence and wavelength bandwidth were performed, as well as with different monochromating crystals or sources, the results of which will also be discussed.  For the instrumental configuration employing PG as a monochromator and analyser, both the monochromator crystal and analyser were oriented for the (002) Bragg reflection for which the atomic plane spacing is 3.3539 Angstrom.  The nominal beam wavelength was 4.75 Angstrom.  For those measurements in which polarized beams were employed, the polarizing efficiency of the instrument was approximately 0.98 for either spin ``+'' or spin ``-'' neutron eigenstates.  The polarization or quantization axis of the instrument was along the vertical direction, perpendicular to the scattering plane.  The magnetization of the Ni stripes of the gratings was saturated in a field of approximately 200. Gauss (0.02 Tesla) for alignment of the stripe along the field direction.  Once again, little will be said about magnetic effects in this paper other than to point out a splitting in the critical angles for spin + and - neutrons for total external reflection from the Ni film.  Phenomena associated with the interaction of the neutron spinor wave function and the magnetic material of the grating is a subject for another study.

\begin{figure}
\includegraphics[width=\linewidth]{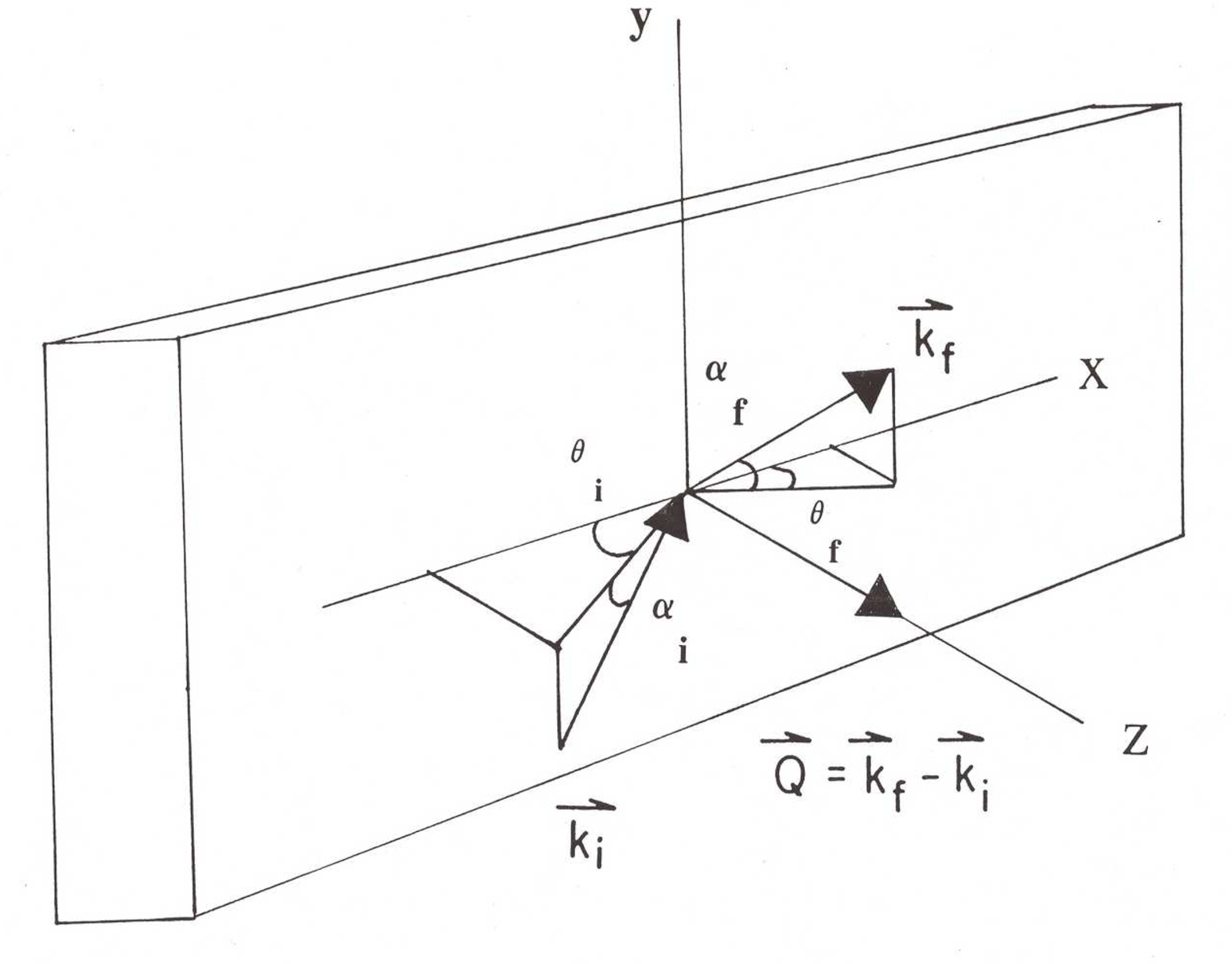}
\caption{\label{fig:5} 
Incident and reflected nominal neutron wavevectors in terms of angles referred to the rectangular coordinate system of the flat sample film and substrate.  (Adapted from Fig. 12.12 of \cite{majkrzak2006structural}.)}
\end{figure}

Figure \ref{fig:5} is a schematic of the incident and elastically reflected nominal neutron wavevectors associated with an individual neutron wave packet referred to the sample coordinate system in terms of angles.  The wave vector transfer $\mathbf Q = \mathbf k_F - \mathbf k_I$  .  Equation \ref{eq:4} gives the relationship between initial $(I)$ and final $(F)$ wave vector components and angles ($k = 2\pi / \lambda$ where $\lambda$ is the neutron wavelength).

\begin{equation}
  \label{eq:4}
  \begin{array}{l l}
  k_{sx} &= k \cos \alpha_s \cos \theta_s \\
  k_{sy} &= k \sin \alpha_s \\
  k_{sz} &= k \cos \alpha_s \sin \theta_s
  \end{array}
\end{equation}                                             
where $s = I$ or $F$).  As mentioned previously, the vertical divergence of the beam is approximately 2.5 degrees FWHM so that the incoherent instrumental $Q$ resolution along the vertical y-axis is relatively coarse.  However, for the specular and non-specular measurements performed here at relatively small angles of glancing incidence relative to the $x$-axis, the $Q$ resolution can be relatively fine and sufficient to resolve real space structures with dimensions of thousands of Angstroms.  Approximate analytic expressions for the instrumental incoherent resolution are given in the following discussion.

Because of the diverse range of optical elements employed within the reflectometer to define both incident and reflected beams, the angular and spatial profiles of the beams are neither simply Gaussian nor trapezoidal.  For example, the monochromator crystal has an angular distribution of perfect microcrystalline mosaic blocks that is approximately Gaussian whereas the pairs of slits up- and down-stream of the sample define a trapezoidal distribution.  However, for sufficiently small aperture widths, from about 0.2 through 1.0 mm, the lineshape closely resembles a Gaussian with a well-defined peak and characteristic full width at half maximum (FWHM) that is approximately twice the length of the base of the distribution.  For all of the measurements of nonspecular reflection described in this paper, the collimations used are within this smaller aperture range.  For the specular reflectivity measurements described herein, both finer and coarser apertures were used.  It is straightforward to apply incoherent instrumental resolution corrections in terms of fractional uncertainties in angle and wavelength (using a formula for $dQ/Q$ given later on).

Relatively simple geometrical or ray tracing formulas can be derived to calculate the FWHM values of angular scans of both the incident beam as a function of scattering angle SA (no sample in beam) and of the so called ``rocking curve'' scan of the sample through an angle theta with the detector at a particular fixed scattering angle (for the case of a perfectly flat substrate -- distortion of the substrate will be dealt with later on).  These formulas are given below in terms of the instrumental slit widths and distances between elements.  These calculations agree with measured widths to an accuracy of about one percent (e.g. for a scattering angle or detector scan with 0.2 mm slit widths, the calculated FWHM is 0.0237 degrees compared to the measured value of 0.0234 -- a difference of about 1.32 \%).  For the scattering angle $SA$, the FWHM is given by

\begin{equation}
  \label{eq:5}
  \underset{\mathrm{(FWHM)}}{\Delta SA}\!\!\! \approx \!\! \arctan (\frac{1}{2 L_F L_I})[(2W_I)(L_S + L_F) + (L_I)(W_I +W_F)]
\end{equation}
where $W_I$ is the slit width of each of the pair of slits upstream of the sample which define the incident beam angular divergence in the scattering plane, $W_F$ is the slit width before the detector, $L_I$ is the distance between the upstream pair of slits (1549. mm), $L_S$ is the distance between the sample center and the closest upstream slit (356. mm), and $L_F$ is the distance between the sample center and the detector slit (864. mm).  A typical value for $W_I$ is 0.2 mm and $W_F$ usually varies between 0.2 and 1.0 mm.

Similarly, the rocking curve FWHM is given by

\begin{equation}
  \label{eq:6}
  \Delta \theta_{RC(\mathrm{FWHM})} \approx (\beta + \gamma)/2
\end{equation}

where
\begin{equation}
  \label{eq:7}
  \beta = \arctan (D / (2L_F))
\end{equation}

\begin{equation}
  \label{eq:8}
  D = ( W_I  /  L_I )[2(L_S  +  L_F) +  L_I]
\end{equation}

and
\begin{equation}
  \label{eq:9}
  \gamma = \arctan ( W_F / (2 L_F))   
\end{equation}

\begin{figure}
\includegraphics[width=\linewidth]{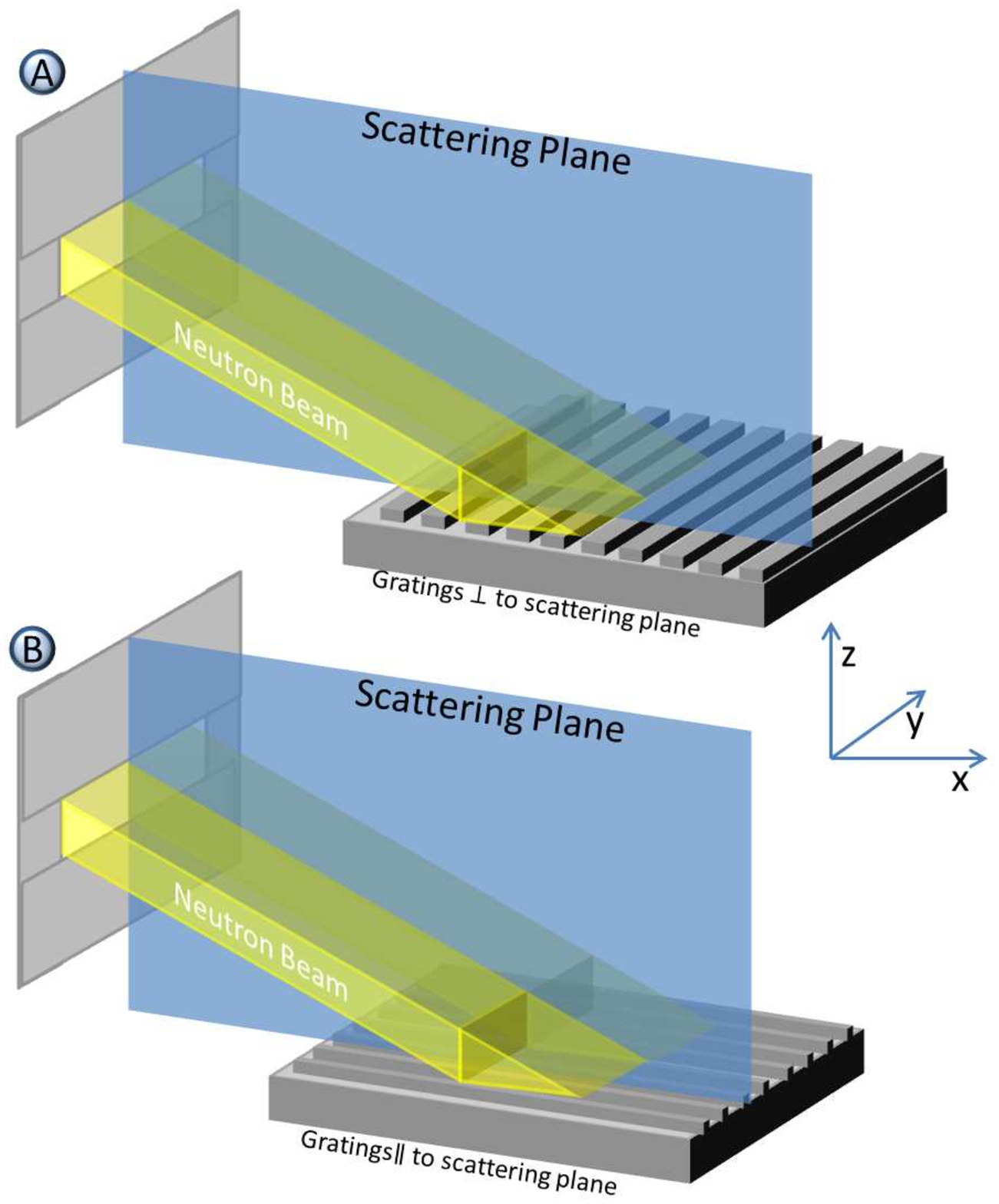}
\caption{\label{fig:6} 
(Color online)  (A) Grating stripes perpendicular to scattering plane defined by incident and reflected nominal wavevectors.  (B)  Grating stripes parallel to scattering plane.  In the actual instrumental configuration, the scattering plane is horizontal -- it is depicted rotated by 90 degrees in the figure for clarity.}
\end{figure}

The conventional instrumental beam properties described above are used in the analysis of the specular and, especially, the non-specular scattering from the grating samples discussed in following sections.
Figures \ref{fig:6}a and \ref{fig:6}b depict the two orientations of the gratings for which reflectivity measurements were performed.

\section{\label{sec:gratings_details}Description, fabrication, and characterization of the diffraction gratings}
Diffraction gratings were constructed of a variety of stripe materials deposited on 3 inch diameter by 1 mm thick Si substrates.  The stripes were nominally of rectangular cross section and 1000 Angstroms thick with the stripe width approximately equal to the spacing between stripes (so that the period of the grating is roughly twice the stripe width).  Stripe materials included both Ni and Permalloy, but most of the measurements reported on here are for the Ni stripe gratings.  Grating periods of 1600 (800 + 800), 800 (400 + 400), 400 (200 + 200), 200 (100 + 100), 100 (50 + 50), 50 (25 + 25), 20 (10 + 10) and 10 (5 + 5) micrometers were fabricated.

Samples were characterized by stylus profilometry, optical microscopy, and scanning electron microscopy. Unpatterned thin films for each deposition instrument were characterized using x-ray reflectometry. All deposition techniques produced films with less than 3\% relative thickness variation across the wafer. All gratings were uniformly spaced and feature sizes showed minimal variation.

The gratings were also characterized by optical diffraction measurements, primarily to establish the intrinsic correlation length of the grating stripes.  An optical reflectometer was constructed with concentric sample and detector arm rotation axes using components comparable in precision and accuracy to those  employed on the neutron reflectometer.  A laser with a nominal wavelength of 655 nm, an angular divergence of about 0.01 degrees, and a spatial width of approximately 2.3 mm in the scattering plane (at the detector aperture located 24. cm from the sample rotation axis) was used as a light source.  The linear detector aperture was placed just in front of the photocell detector with a width of approximately 0.05 mm in the scattering plane (no other apertures were used -- the exit of the laser was about 40 cm from sample center).  (Any diffraction of the light by the detector aperture, after reflection from the sample grating, was integrated over by the relatively wide photocell behind and nearly adjacent to the slit.)

Without any sample, a detector (scattering angle) scan yielded an effective instrumental angular width of $\approx 0.25$ degrees.  A $Q_x$ scan at a constant value of $Q_z = 0.0016 \AA{}^{-1}$ was performed.  The widths of the principal reflections (m = +/- 1), higher orders, and specular ridge (m = 0) were identical, to within experimental uncertainty at a value in sample angle of approximately 0.125 degrees FWHM -- in other words, limited by the instrumental resolution.  It is of particular note that the widths of the higher order reflections did not display an increasing intrinsic width with increasing Q -- which would have been indicative of a loss of correlation due to static disorder in the periodicity of the grating unit repeat structure.  The uncertainty in the line width measurement is of the order of 10 \% (upper limit).  If we conservatively attribute this amount to a natural correlation length, then deconvoluting this amount from the measured width $(3.8 \times 10^{-6} \AA{}^{-1})$ of the $m = 1$ reflection gives a minimum correlation length for the grating of about 376. microns.  This is a value significantly greater than the transverse coherence length of the neutron wave packet we are attempting to measure, even appreciably more than its corresponding projection on the surface at the low angles of reflection where the measurements were performed.  Thus the crucial assumption underlying the use of the diffraction grating structure as a measuring stick of the neutron transverse coherence length, namely that the grating period correlation length be sufficiently large in comparison, is well justified.

\section{\label{sec:specular_analysis}Analysis of the specular reflectivity}
Specular neutron reflectivity measurements with the wavevector transfer perpendicular to the surface of the sample film and substrate provide information about the scattering length density (SLD) depth profile along the surface normal \cite{majkrzak2006structural} with spatial resolutions down to about half a nanometer possible (under proper conditions).  Such measurements have become established over the last two decades or so as a valuable probe of thin film structure for materials of interest in physics, chemistry, and biology \cite{kirby2012phase}.  In the absence of any significant in-plane variations in SLD, the analysis of specular neutron reflectivity data to extract SLD profiles reduces to a one-dimensional problem involving a piece-wise continuous solution of the time-independent Schroedinger wave equaton.  It is even possible, using phase-sensitive techniques involving adjacent reference structures, to obtain the complex reflection amplitude containing all of the phase information and subsequently perform an exact, first-principles inversion for the SLD profile -- which is unique to within limits determined only by statistical noise and truncation of the data at some practical maximum value of $Q$ \cite{majkrzak2003phase}.  For a perfectly smooth and flat semi-infinte substrate of uniform SLD (and with no absorption), the reflectivity is unity below a critical value of $Q$ given by

\begin{equation}
  \label{eq:10}
  Q_c^2 = 16\pi\rho
\end{equation}
Above this critical $Q_c$, the reflectivity falls off as the inverse fourth power of $Q$.

If in-plane variations in SLD are present but of sufficiently small magnitude and/or spatial extent that the specular analysis remains a good approximation, then the effective SLD depth profile corresponds to the in-plane average SLD as defined by \cite{majkrzak2006structural}

\begin{equation}
  \label{eq:11}
  r = \frac{4\pi}{iQ} \int_{-\infty}^{\infty} \psi_{kz}(z) \langle\rho(x,y,z)\rangle_{xy} e^{ikz} ~dz
\end{equation}
where the value of $k$ in the argument of the exponential is $k = k_{0z}$ and where

\begin{equation}
  \label{eq:11a}
  \langle\rho(x,y,z)\rangle_{x,y} = \frac{1}{A}\iint_{-\infty}^{\infty} \rho(x,y,z) ~dx~dy
\end{equation}
and A is the in-plane area.  If, on the other hand, the in-plane SLD variations are large enough, the one-dimensional approximation for the treatment of the specular reflectivity is no longer valid and the 3-dimensional scattering problem must be solved.  This is a considerably more formidable task, but one that we must ultimately deal with below (and in the accompanying article, Part II) in a quantitative analysis of the specular reflectivity for some of the gratings with periods that fall within an intermediate or cross-over region.

Nonetheless, it turns out that for the geometry where the grating stripes are parallel to the scattering plane, and for certain asymptotic cases in the limit of either larger or smaller grating periods with stripes perpendicular to the scattering plane, it suffices to observe the overall shape of the reflection curve between the critical angles, for total external reflection, of Ni and Si.  The particular analysis of the neutron transverse coherence length which we adopt here is based on that discussed in \cite{majkrzak2006structural}.  This approach has recently been applied to x-rays \cite{lee2011determining}.  Other determinations of transverse coherence employing grating structures have been performed, e.g., by Salditt et al. \cite{salditt1994x}, but by analyzing the widths of off-specular grating reflections.

\begin{figure}
\includegraphics[width=\linewidth]{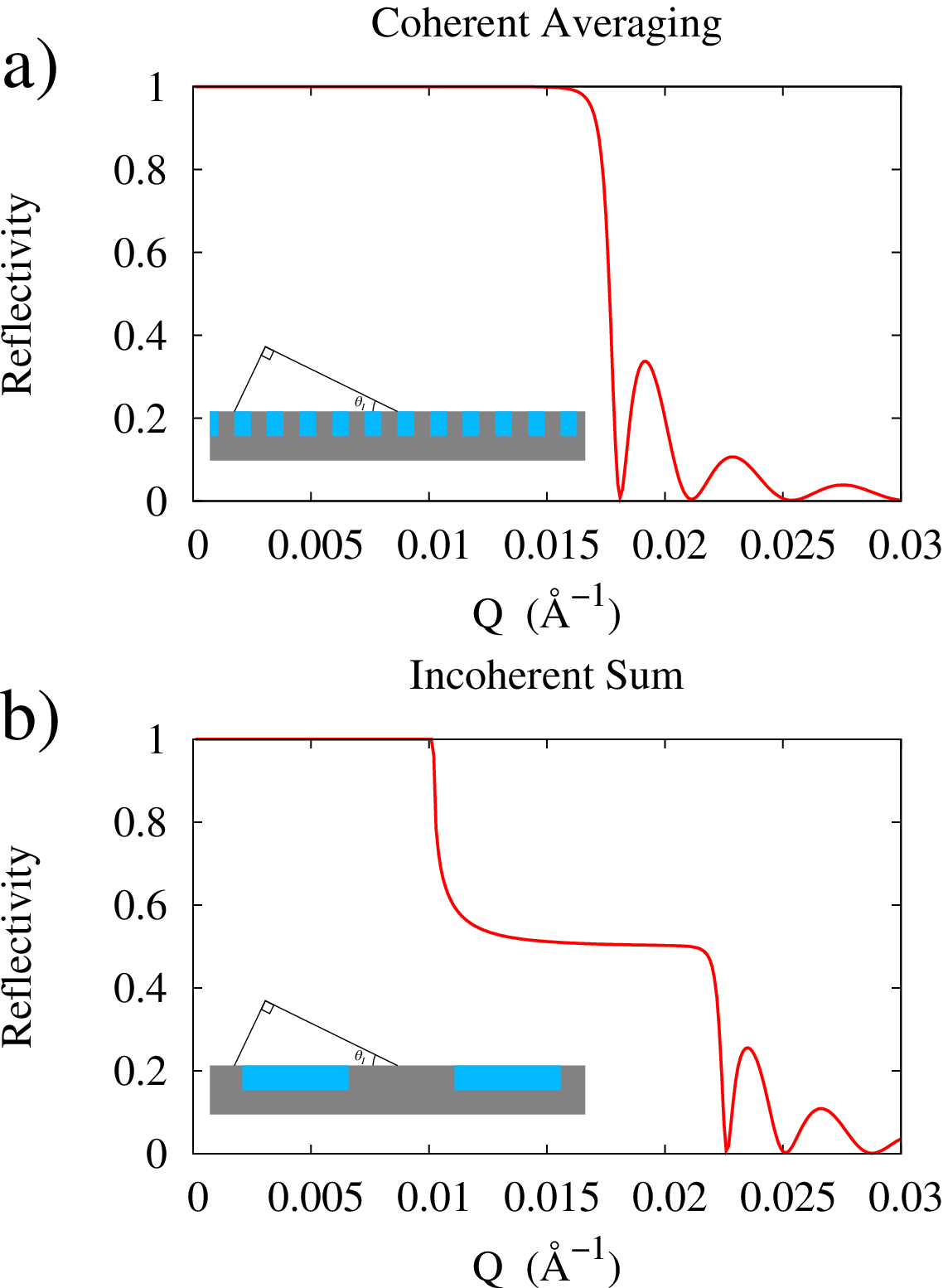}
\caption{\label{fig:7} 
(Color online)  a) Reflectivity curve corresponding to an effective coherent averaging of two different SLDs, as described in the text.  In the real space schematic of the grating structure in the inset, the material for the periodic rectangular structure is the same as the substrate and is taken to have the SLD of Si -- the troughs in between, on the other hand, are filled with material having the SLD of ordinary Ni.  Only a single critical Q is observed.  Note that the real space figure in the inset is a highly schematic representation of the projection of the wave packet's transverse coherence length onto the grating.  In reality, the wavefront structure of the packet is highly distorted in the vicinity of its interaction with the material as can be shown by basic one-dimensional calculations of a simple plane wave incident upon a potential barrier -- and which accurately describes specular neutron reflection.  (See, for example, \cite{cohenlaloe} or the article by Goldberg et al. \cite{goldberg1967edge}.) (Adapted from Fig. 12.11a of \cite{majkrzak2006structural}.)
b) Reflectivity curve corresponding to an incoherent sum of two independent areas of in-plane SLD in the film, as described in the text.  Note the appearance of two distinct critical Q values.(Adapted from Fig. 12.11b of \cite{majkrzak2006structural}.)}
\end{figure}

A typical specular reflectivity curve for a film (approximately 1000 Angstroms thick) composed of equal areas of two different constant SLDs (for Ni and Si on a semi-infinite substrate of Si) --  and for which the dimensions of the in-plane areas of each of the two materials are sufficiently smaller than the projected length of the transverse dimension of the neutron wave packet (as depicted in the inset) --  is shown in Figure \ref{fig:7}a.  In this case a single critical $Q$ is observed which is related to a SLD that is the average of that for the two materials (as follows from Equation 11).  In Figure \ref{fig:7}b is another specular reflectivity curve for a film which has, again, equal net areas of the same two different constant SLDs but where now the areas of each material are divided into constituent sub-areas which have dimensions large enough that effective coherent averaging does not appreciably occur.   In this latter case, the observed reflectivity is essentially an area-weighted incoherent sum of two distinct reflectivities, each for one of the two different SLDs.  In this latter case, two critical $Q$ values are observed, one corresponding to the material of lower SLD (Si) (below which the reflectivity is unity) and another determined by the SLD of the material of higher SLD (Ni).  Between these two critical $Q$ values the reflectivity is 50 percent, corresponding to the equal in-plane areas of the two materials.  The differences between the reflectivity curves for the two cases shown in Figures \ref{fig:7}a and \ref{fig:7}b are striking and clearly indicative of the length scale over which effective averaging takes place.  Moreover, the incoherent instrumental $Q$-resolution associated with the beam (as discussed earlier) has only a  relatively weak affect on these specular curves in the vicinity of the critical values of $Q$, as will be demonstrated below.

Table \ref{table:1} lists values of the critical wavevector $Q_c$ for a number of different materials relevant to the present work.
\begin{table}
\caption{\label{table:1}Values of the Critical Wavevector $Q_c$ for Relevant Materials}
\begin{ruledtabular}
\begin{tabular}{ll}
Material & $Q_c (\AA {}^{-1})$ \\
\hline
Ni (unmagnetized) &  0.0217 \\
Si &  0.0102 \\
50 \% Ni + 50 \% Si (by volume) & 0.0170 \\
50 \% Ni + 50 \% vacuum (by volume) & 0.0154 \\
\end{tabular}
\end{ruledtabular}
\end{table}

The incoherent instrumental resolution along the z-axis for specular reflection, $dQ/Q$, results from angular beam divergence and wavelength spread and is given by
\begin{equation}
  \label{eq:12}
  \Delta Q_z / Q_z \approx [(\Delta \lambda / \lambda)^2 + (\Delta \theta / \theta)^2]^{1/2}
\end{equation}
Typical values of $dQ/Q$ (e.g., 0.025) affect the specular reflectivity curves primarily by a relatively mild rounding near the critical $Q$ values and broadening of the interference fringes, due to the overall film thickness, just above the critical $Q$.  (The incoherent instrumental $Q$-resolution along the grating substrate surface normal is not significantly affected by the angle α, as defined in Figure \ref{fig:5} and Equation (\ref{eq:4}), since its maximum magnitude is no greater than approximately 2.5 degrees, as mentioned previously.  However, the ability to relax the angular divergence along the y-axis perpendicular to the nominal scattering plane, using the slit geometry, results in a significant enhancement of useful beam intensity.)

The incoherent instrumental resolution, therefore, does not significantly interfere with our ability to determine -- from the specular reflectivity -- the point at which the transverse extent of the coherent volume of the neutron wave packet is capable of effectively averaging over areas of different SLD of a certain size .  Note that for glancing angles of incidence, where the direction of the incident neutron wavevector is close to being parallel to the in-plane x-axis, the projected length of the transverse coherence length is greatly enhanced.  Thus, for grating stripes oriented perpendicular to the scattering plane (as in Figure \ref{fig:6}a), a given transverse neutron coherence length can simultaneously interact with a larger number of grating stripes.

\subsection{\label{subsec:pyrolytic_graphite_monochromator}Ideally imperfect, pyrolytic graphite mosaic crystal monochromator}
\subsubsection{Grating stripes parallel to the scattering plane}
\begin{figure*}
\includegraphics[width=\linewidth]{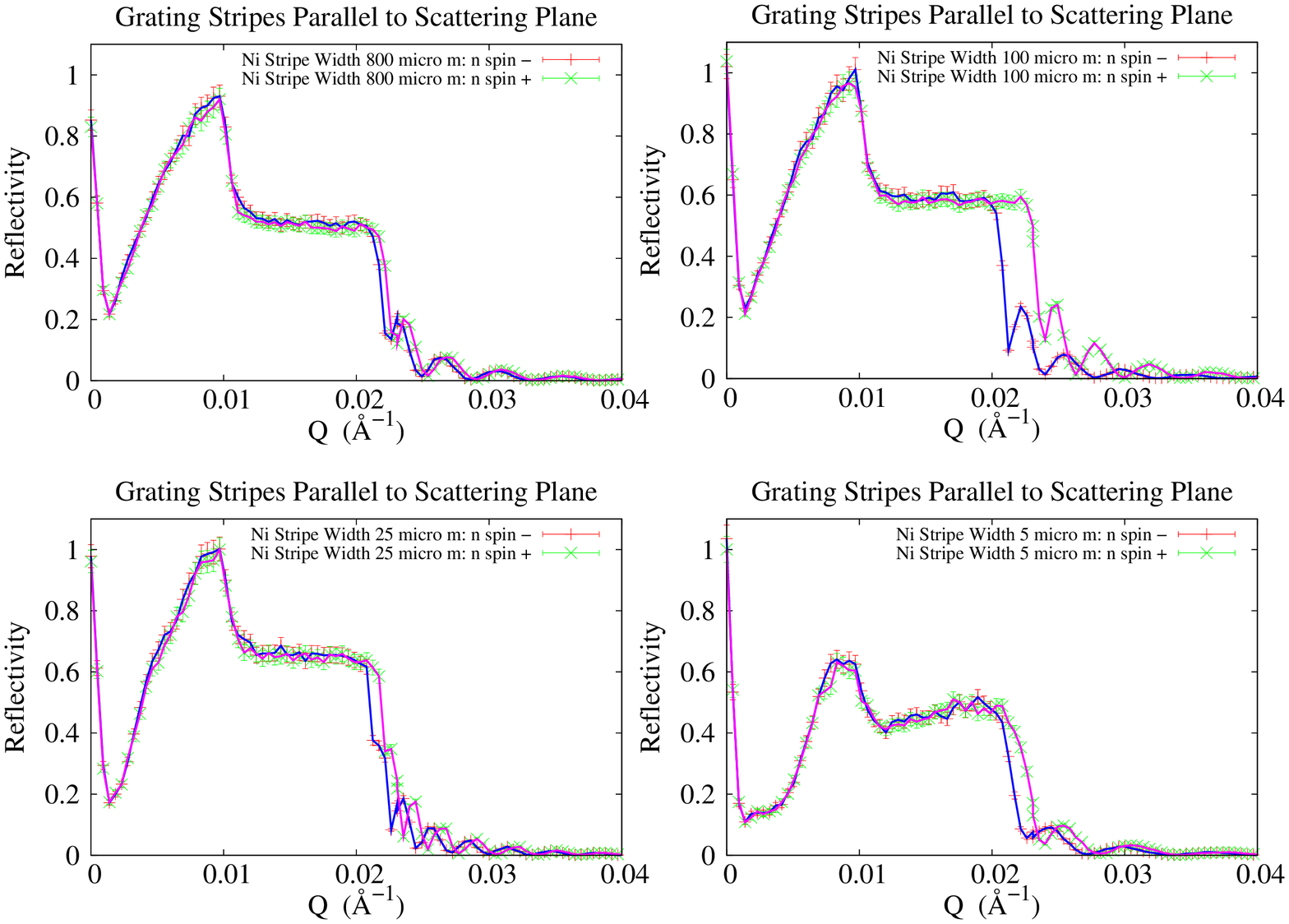}
\caption{\label{fig:8} 
(Color online)  Measured specular reflectivity curves for gratings with stripes parallel to the scattering plane as a function of Ni stripe width.  In all cases measured, including those selected for the figure, the specular relectivities from Ni and Si areas in plane add independently as an incoherent sum as discussed in the text.  The critical Q values for Si and Ni (unmagnetized) are, respectively, 0.0102 and 0.0217 inverse Angstroms.  Note that the geometrical footprint of the beam on the sample surface is evident below the critical Q for Si and has not been corrected for -- with correction, the reflectivity is essentially unity below this value.  Between the critical angles for Si and Ni, the reflectivity should be 50 \% for equal stripe and trough widths -- that it is somewhat higher than 50 \% for some samples may be indicative of a somewhat larger stripe width.  The degree in splitting at the Ni critical edge depends on the saturation of the Ni film magnetization in an applied field of only about 200 Gauss (0.02 Tesla).}
\end{figure*}

For all of the measurements performed on the gratings oriented with the stripes parallel to the scattering plane, from periods of 1600 all the way down to 10 micrometers, the measured reflectivity curve was found to represent an incoherent sum of reflected intensities corresponding to the two different material SLDs.  Note that in this orientation there is no angle-enhanced projection of the transverse coherence length along the direction of the grating modulation as occurs in the orthogonal alignment of the grating. Some of the actual reflectivity curves for parallel stripes of different widths are shown in Figure \ref{fig:8}. This establishes that, along the $y$-direction at least, the neutron transverse coherence length is no greater than of the order of 10 micrometers. 

\subsubsection{Grating stripes perpendicular to the scattering plane}
\begin{figure*}
\includegraphics[width=\linewidth]{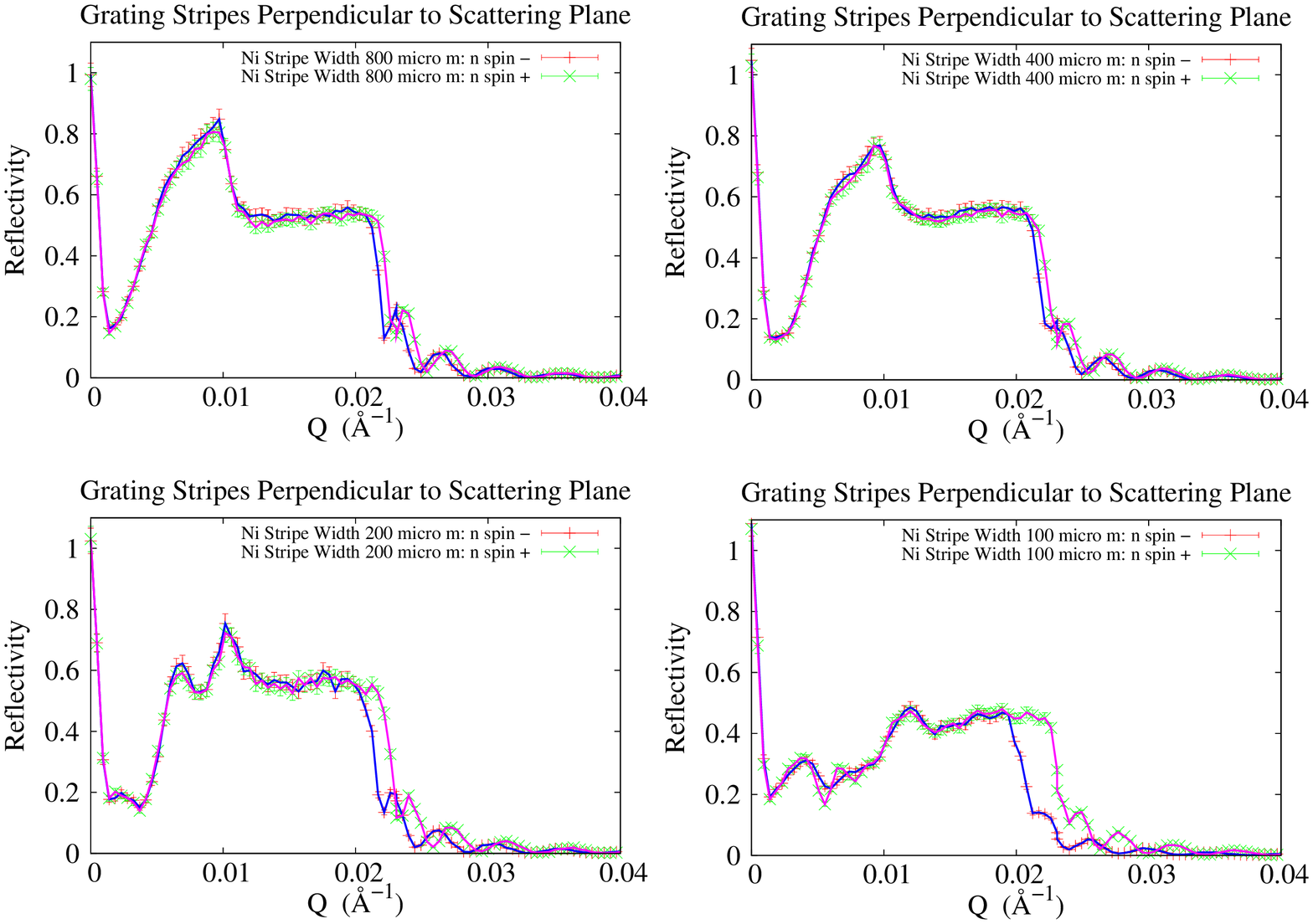}
\caption{\label{fig:9} 
 (Color online)  Grating stripes perpendicular to the scattering plane -- incoherent limit for larger grating periods.  As for the case with the Ni grating stripes horizontal, both Si and Ni critical Qs are observed, indicating an incoherent sum of two independent reflectivities.}
\end{figure*}

In the grating orientation of Figure \ref{fig:6}a, with the enhanced projection of the neutron's transverse coherence, markedly different behavior is observed for sufficiently small grating periods, unlike in the parallel stripe orientation.  Figure \ref{fig:9} shows reflectivity curves for some of the larger grating periods, down to about 400 micrometers, where the measured resultant reflectivity is clearly still an incoherent sum of the reflectivities associated with the Ni film and underlying Si substrate.  Note that in between the nominally 1000 Angstrom thick Ni film stripes is vacuum -- i.e., in the gratings fabricated for the measurements of this study, the troughs are not filled with Si as in the hypothetical case depicted in Figure \ref{fig:7}.

\begin{figure}
\includegraphics[width=\linewidth]{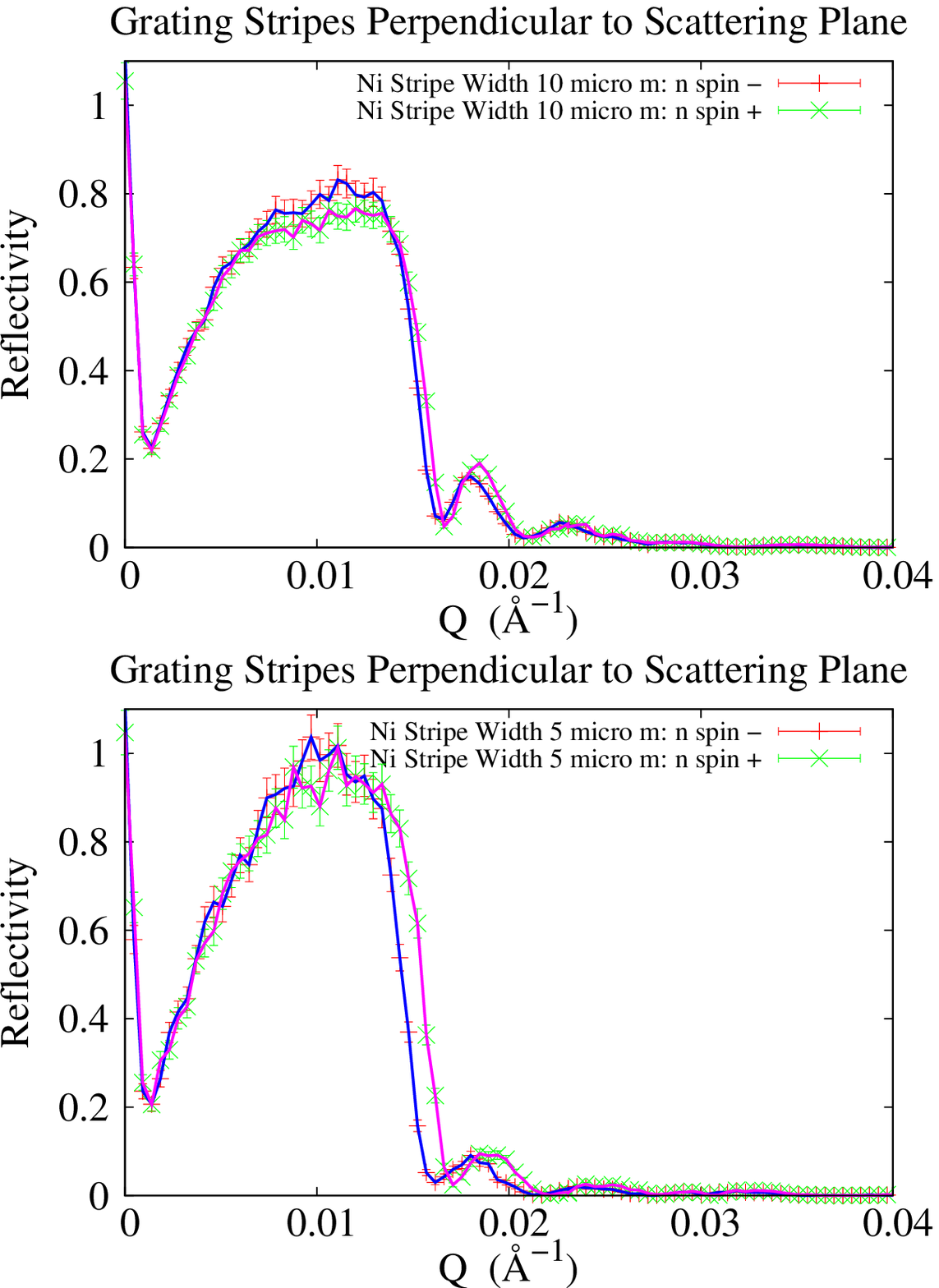}
\caption{\label{fig:10} 
 (Color online)  Grating stripes perpendicular to the scattering plane -- coherent limit for smaller grating periods.  There is now only a single critical $Q$, that corresponding to a film with half the SLD of Ni on a semi-infinite Si substrate.  The neutron wave packet has a transverse coherence sufficient to effectively average over multiple grating periods to produce a measured specular reflectivity that corresponds to a coherent in-plane average of the SLD of grating stripe plus empty trough.}
\end{figure}

In contrast to this result, Figure \ref{fig:10} shows reflectivity curves for gratings with spacings of 20 and 10 micrometers.  In this limit, the neutron wave packet effectively averages over a sufficient number of periods in the grating structure.  There is now only a single critical $Q$ -- that corresponding to a film with half the SLD of Ni on a semi-infinite Si substrate.  The neutron wave packet has a transverse coherence sufficient to effectively average over multiple grating periods to produce a measured specular reflectivity that corresponds to a coherent in-plane average of the SLD of grating stripe (Ni) plus trough (empty).

\begin{figure}
\includegraphics[width=\linewidth]{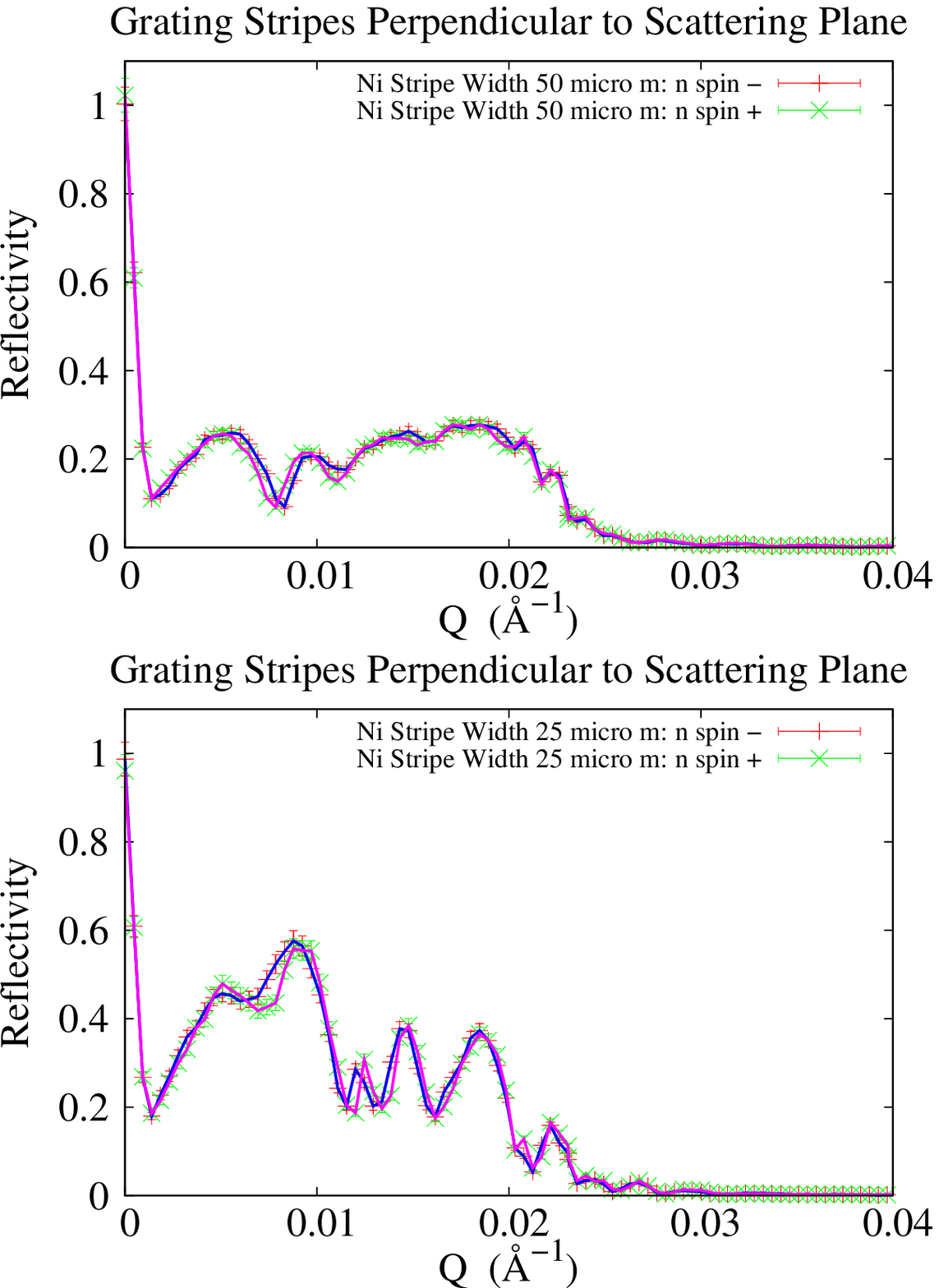}
\caption{\label{fig:11} 
(Color online)  Specular reflectivity for grating periods in an intermediate regime between completely coherent and incoherent limits where a much more complicated specular reflectivity curve is observed.  These relatively large distortions are also a manifestation of the grating structure itself, as discussed in the theory section and associated theory paper (Part II).  Moreover, non-specular scattering can occur simultaneously with the specular process and, depending on the aperture width in front of the detector, may or may not be distinguishable.}
\end{figure}

In between these two limiting regimes, however, for grating periods of 50 and 100 micro meters, a much more complicated specular reflectivity curve is observed, as shown in Figure \ref{fig:11}, where a cross-over region exists between that in which effective in-plane coherent averaging occurs and one for which sufficiently large in-plane areas of different SLD contribute to an incoherent sum of independent reflected intensities.  These relatively large distortions observed in the specular reflectivity for the intermediate grating periods are also a manifestation of the grating structure itself, as discussed in the theory section and associated theory paper (Part II).  Moreover, non-specular scattering can occur simultaneously with the specular process and, depending on the aperture width in front of the detector, may or may not be distinguishable.

\begin{figure}
\includegraphics[width=\linewidth]{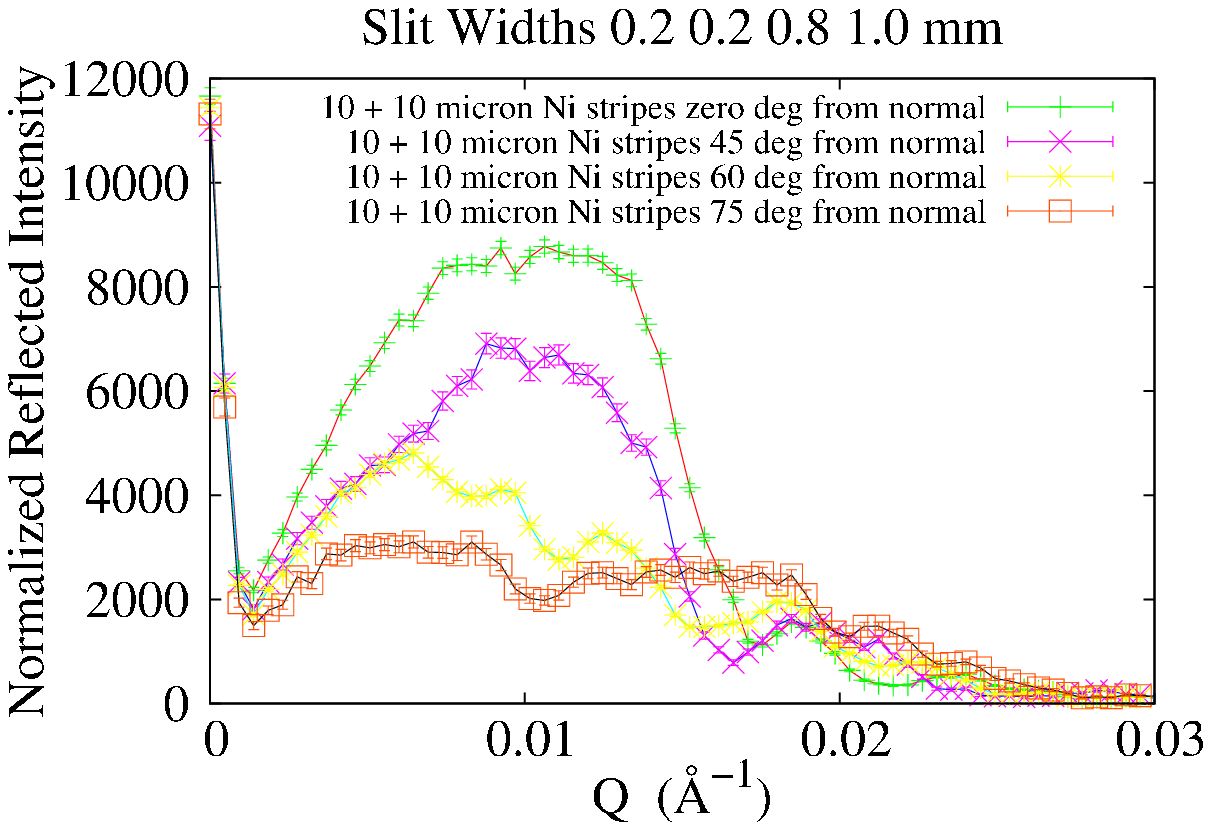}
\caption{\label{fig:12} 
(Color online)  Specular reflectivity as a function of rotation angle of the Ni stripes away from the vertical orientation (which is normal to the scattering plane defined by incident and reflected wavevectors).  Note that going from a value of Q = 0.01 to Q = 0.02 inverse Angstroms (along the horizontal axis) corresponds, approximately, to the in-plane projection $L_x$ of a 1 micron (for instance) neutron transverse coherence $\Delta r_\perp$ decreasing from 264. to 132. microns (see Equation \ref{eq:20}).  The effective grating periods at 0.0, 45., 60., and 75. degrees from vertical (corresponding to the curves in the plot, in order, from uppermost to lowest) are, respectively, 20., 28., 40., and 77. microns.  In the above graph, the appearance of a critical edge near that corresponding to Ni (0.0217 inverse Angstroms) begins to emerge at an effective projection of 40. microns, indicating an incoherent condition.  At this $Q$ value, the angle of incidence is about 0.47 degrees (for a wavelength of 4.75 Angstroms) so that the transverse coherence length of the packet would be roughly 0.33 microns.   However, as discussed in the text, a more accurate estimate of the coherence length must be made through a proper quantitative analysis of the entire specular reflection curve and off-specular grating reflection widths -- which is done in subsequent sections.}
\end{figure}

Because no grating periods between 20 and 50 micrometers were available, we performed specular reflectivity measurements on the 20 micrometer grating where the stripes were rotated away from the vertical orientation along the $y$-axis (stripes nominally perpendicular to incident neutron wavevector).  By so doing, it was possible to effectively tune the grating period continuously so that a better measure of the transverse coherence length could be obtained.  A more accurate determination of the coherence length can be made, in principle, through a proper quantitative analysis of the entire specular reflection curve and off-specular grating reflection widths.  (We discuss such calculations in a subsequent section and, in more detail, in the associated paper, Part II.)  Nonetheless, a semi-quantitative estimate can be made, keeping in mind, for instance that in going from a value of $Q = 0.01$ to $Q = 0.02$ inverse Angstroms (along the horizontal axis of Figure \ref{fig:12}) corresponds, approximately, to the in-plane projection $L_x$ of a 1 micron (for example) neutron transverse coherence $\Delta r_\perp$ decreasing from 264. to 132. microns (see Equation \ref{eq:20}).  
A critical edge for Ni (at $Q = 0.0217 \AA{}^{-1}$) begins to emerge at a rotation of approximately 60 degrees from nominal normal to the incident neutron wavevector for the 10 + 10 = 20 micron grating.  The first appearance of such a critical edge would coincide with an effective projected transverse coherence length that was just insufficient to coherently average over the grating structure.  (Conversely, at zero degree rotation the \textit{absence} of a Ni critical edge is indicative of effective in-plane averaging.)  At this rotation angle the effective grating period is $20/\cos(60 \mathrm{deg}) = 40$ microns.  The neutron glancing incidence angle is 0.47 degrees (for a wavelength of 4.75 \AA) at this critical value of $Q$ so that $\Delta r_\perp \approx \sin(0.47 \mathrm{deg}) = 0.33$ microns.  The more quantitative analyses of the coherence length from both the specular and nonspecular measurements that follow below yield a value of $\Delta r_\perp$ closer to about one micron.  It will be useful in subsequent discussion of wave packet coherence versus incoherent beam resolution to note that the results illustrated in Figure \ref{fig:12} can be viewed in an alternative way -- the 10 + 10 = 20 micron period grating is coherently averaged over by the neutron wave packet prepared in this instrument whereas a grating of twice the period (the tilted equivalent to a 20 + 20 = 40 micron period) is not.

\begin{figure}
\includegraphics[width=\linewidth]{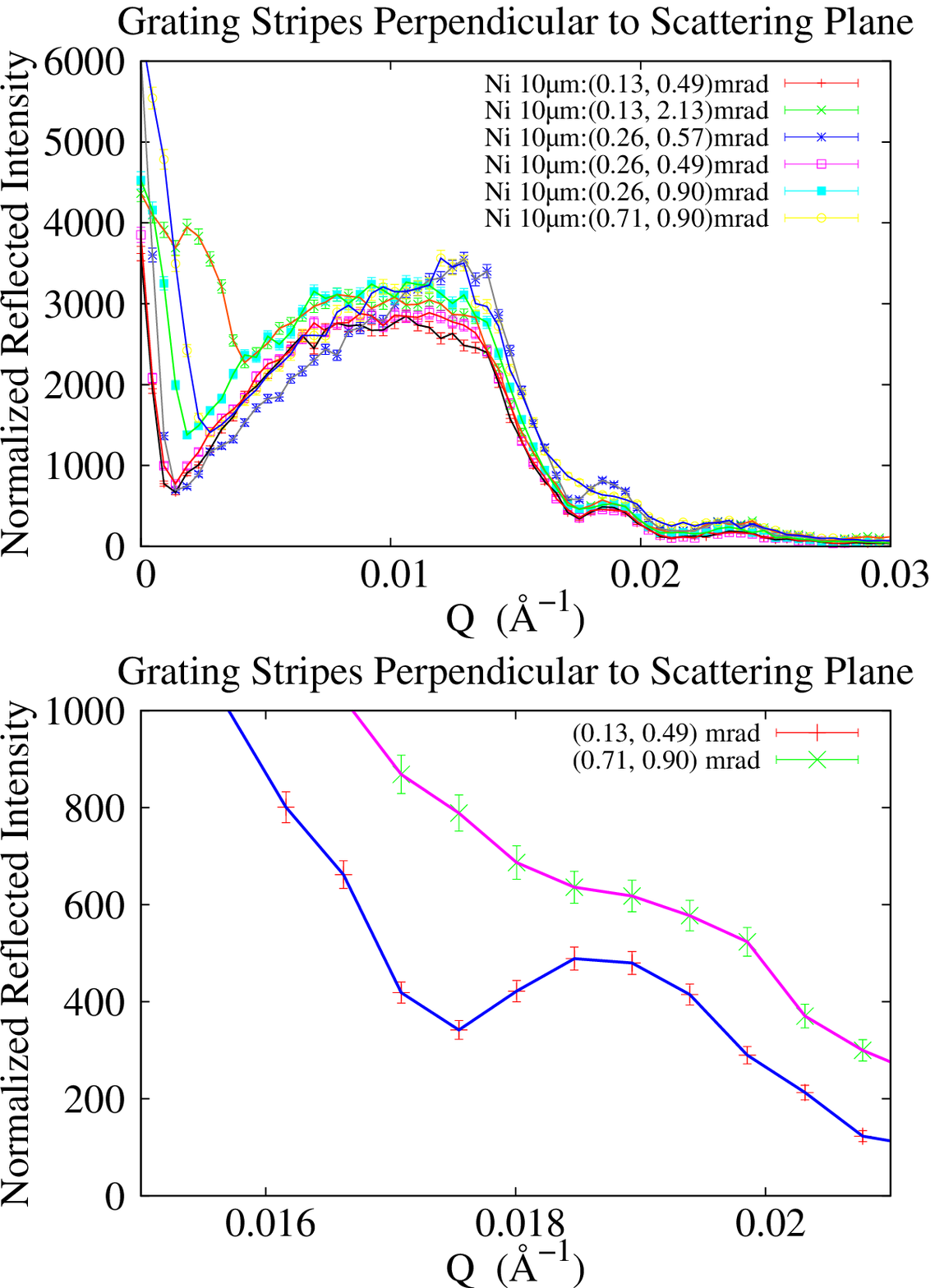}
\caption{\label{fig:13} 
 (Color online)  Upper part -- specular reflectivity as a function of the incoherent instrumental (beam) resolution (the angular divergences for each measurement are indicated in the legend of the figure and differ from tightest to coarsest by a factor of 5 -- the first number is the beam angular divergence in mrad as defined by a pair of slits upstream of the sample whereas the second number pertains to the divergence of the reflected beam) for a grating period (10 + 10 = 20 micron) in the coherent limit of effective averaging by the neutron wave packet.  The incoherent instrumental resolution associated with a beam's angular divergence, where the beam is taken to be composed of an ensemble of nearly identical neutron wave packets having a distribution of nominal wavevector directions, does not have any significant effect on the manifestation of coherent averaging by an individual packet in the specular reflectivity curve.  Note, however, that the beam resolution $\Delta Q$ (which depends primarily on the beam angular divergence) does have a significant effect on the resolution of the interference fringes associated with the 1000 $\AA$ thickness of the grating's Ni stripes -- the first fringe is shown on the bottom as a corresponding detail (of the main figure on the top) for the finest and coarsest beam angular divergences.}
\end{figure}

\begin{figure}
\includegraphics[width=\linewidth]{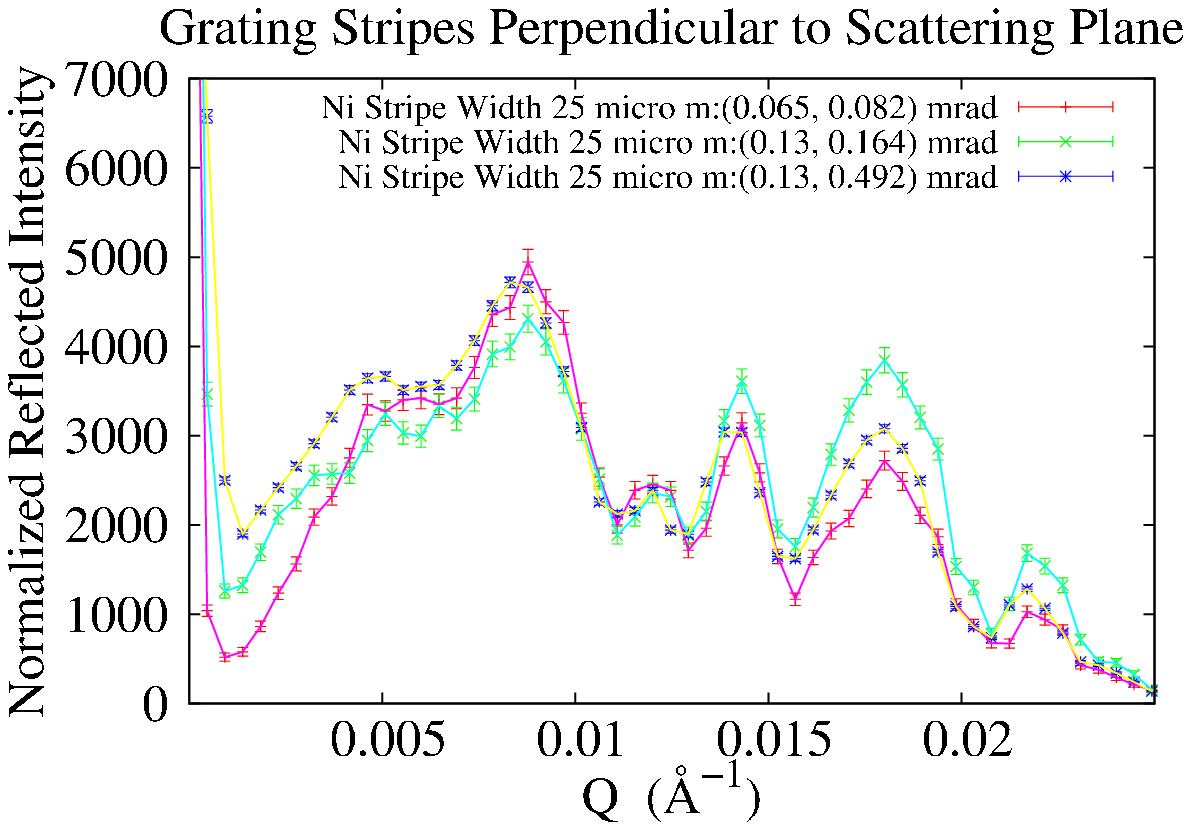}
\caption{\label{fig:14} 
(Color online)  Specular reflectivity as a function of the incoherent instrumental (beam) resolution for a grating period in the intermediate region.  Note the insensitivity of the general shape and features of the specular reflectivity to significant variations in the incoherent instrumental resolution.}
\end{figure}

It is true that the grating structures are 3-dimensional and that non-specular reflection occurs simultaneously with the specular.  Whether specular and nonspecular scattering are collected simultaneously or not in the detector in the course of a specular (theta : two theta scan) for a given grating structure depends on the slit widths and grating period.  These effects must be accurately taken into account in any analysis of the specular component of the reflectivity.  Nonetheless, no pronounced distortions of the specular reflectivity are observed either for the larger or the smaller grating periods in the limiting cases.  Figures \ref{fig:13} and \ref{fig:14} show specular reflectivty data for two different grating spacings, one in the coherent limit and the other in the intermediate regime (10 + 10 = 20 and 25 + 25 = 50 micron periods), as a function of the incoherent instrumental beam divergences.  
The relatively large changes in the incoherent instrumental Q resolution (a factor of 5 from tightest to coarsest) have little effect on the general shapes of the curves and, consequently, on any conclusion drawn regarding the ability of the neutron packet to coherently average (or not).  Recall from the analysis of the results presented in Figure \ref{fig:12} above that changing the grating period by a factor of two (from 20 to 40 microns) is all that is necessary to create an incoherent condition.  Further measurements with even coarser beam resolutions using a vanadium source still show that the wave packet coherence is sufficient to effectively average as shown in a following section.  The present measurement technique makes it possible to effectively separate coherent and incoherent contributions to the effective resolution of the measurement, corresponding to the transverse coherence of the individual neutron wave packet and angular divergences of the beams in the instrument, respectively.

\begin{figure}
\includegraphics[width=\linewidth]{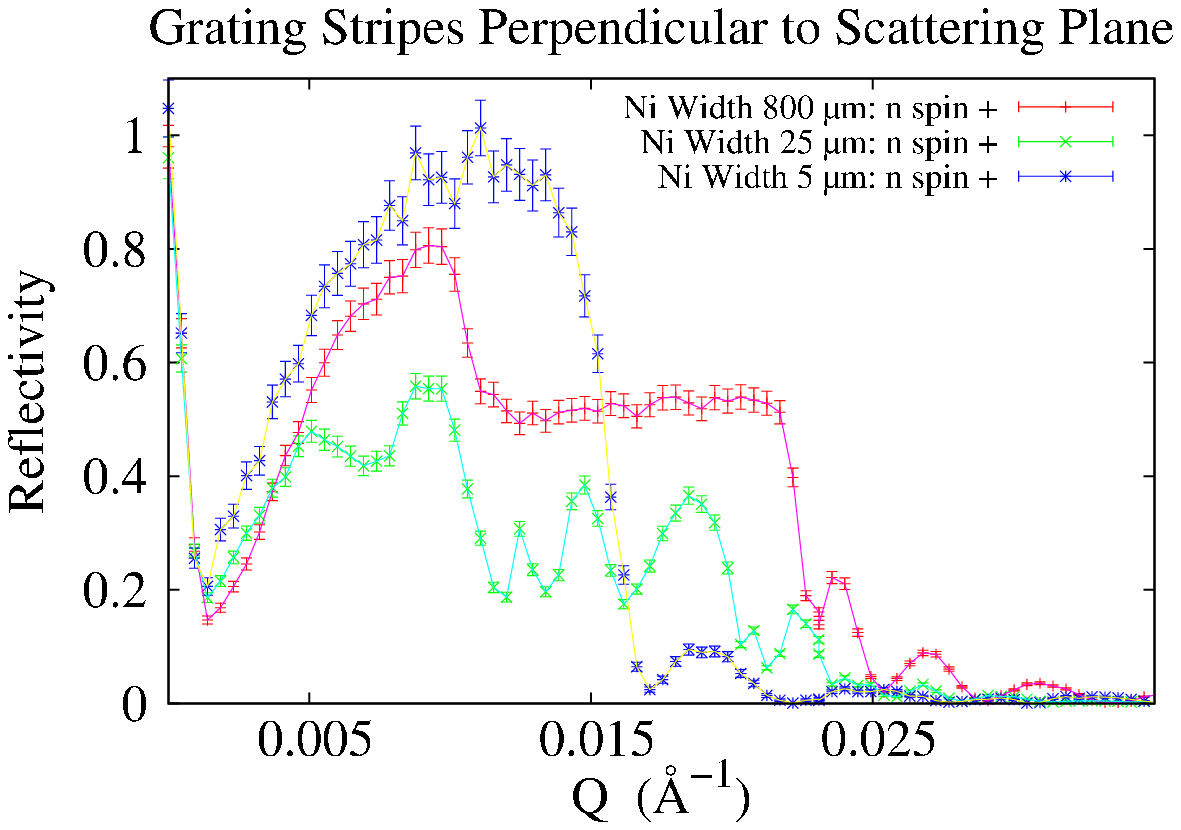}
\caption{\label{fig:15} 
(Color online)  Summary of the specular reflectivity results discussed thus far, showing the reflectivity for a representative grating period in each of the three regions -- incoherent, intermediate, and coherent.}
\end{figure}

Figure \ref{fig:15} summarizes the specular reflectivity results discussed thus far (for measurements in which the mosaic pyrolytic graphite monochromator was used), showing reflectivities for a representative grating period in each of the three regions -- incoherent, intermediate, and coherent.

In a section below, as well as the accompanying paper, Part II, a 2-dimensional wave equation is solved for a model packet wave function which explicitly takes into account the periodic grating structure.  However, before discussing the results of the aforementioned calculation, we present the results of
specular measurements performed with a perfect Si crystal monochromator as well as a spin-incoherent vanadium source (instead of the graphite) in addition to non-specular measurements in which the  transverse coherence length is deduced from an analysis of the line widths of the first-order grating reflections of the 10 + 10 = 20 micron grating.  All of these subsequent measurements and analysis are consistent with an effective transverse coherence of the order of one micron, with relatively little dependence on the beam angular divergence and wavelength bandwidth.

\subsection{Silicon perfect single crystal monochromator}
\begin{figure}
\includegraphics[width=\linewidth]{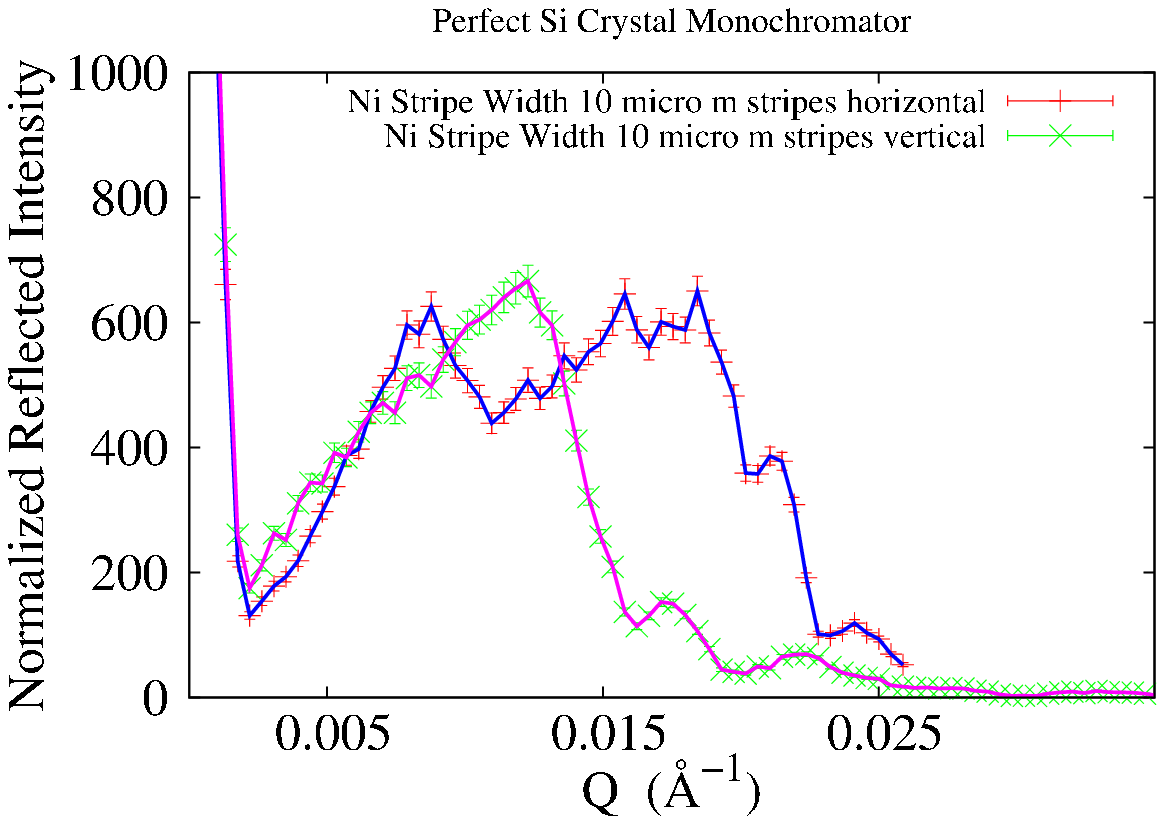}
\caption{\label{fig:16} 
 (Color online)  Selected specular reflection curves obtained with a perfect single crystalline Si monochromator, as discussed in the text.  The results are similar to those obtained with the PG monochromator.  Both the coherent limit -- where in-plane averaging over stripe and trough occurs (vertical stripe orientation) --  as well as the incoherent extreme -- where the reflected beam consists of a sum of two intensities, one from the stripes and the other from the troughs (horizontal stripe orientation) -- are observed.  At the maxima, the reflected intensity is approximately 40 counts/s -- thus, consecutive neutrons traversing the reflectometer are spaced roughly 19 meters apart, on average, whereas the distance between monochromator and detector is only about 3 meters.  This low rate further corroborates the one-neutron-at-a-time nature of the diffraction process. }
\end{figure}

To investigate the degree to which the physical size of a perfect microcrystalline mosaic block might determine the transverse coherence length of a neutron wave packet created within the reflectometer, selected measurements on some of the same grating structures were repeated with a perfect single crystal monochromator instead of the pyrolytic graphite.  The (111) reflection from perfect single crystal silicon was chosen, not only because of its high degree of perfection over macroscopic areas, but also because, at the same angular settings of the instrument used with the PG(002) monochromator, the wavelength is approximately the same (the interatomic plane spacing for PG(002) is 3.354 $\AA$ and for Si(111) it is 3.135 $\AA$).  Figure \ref{fig:16} shows the specular reflection curves obtained with perfect single crystalline Si as monochromator.

\subsection{Incoherent vanadium source}

In place of a crystal monochromator, a pure vanadium cylinder was placed at the monochromator position in the guide cut and used as a source of neutrons for diffraction from the gratings.  Vanadium is an almost completely spin-incoherent scatterer and should, therefore, provide neutrons originating predominantly in the form of isotropic spherical waves.

\begin{figure}
\includegraphics[width=\linewidth]{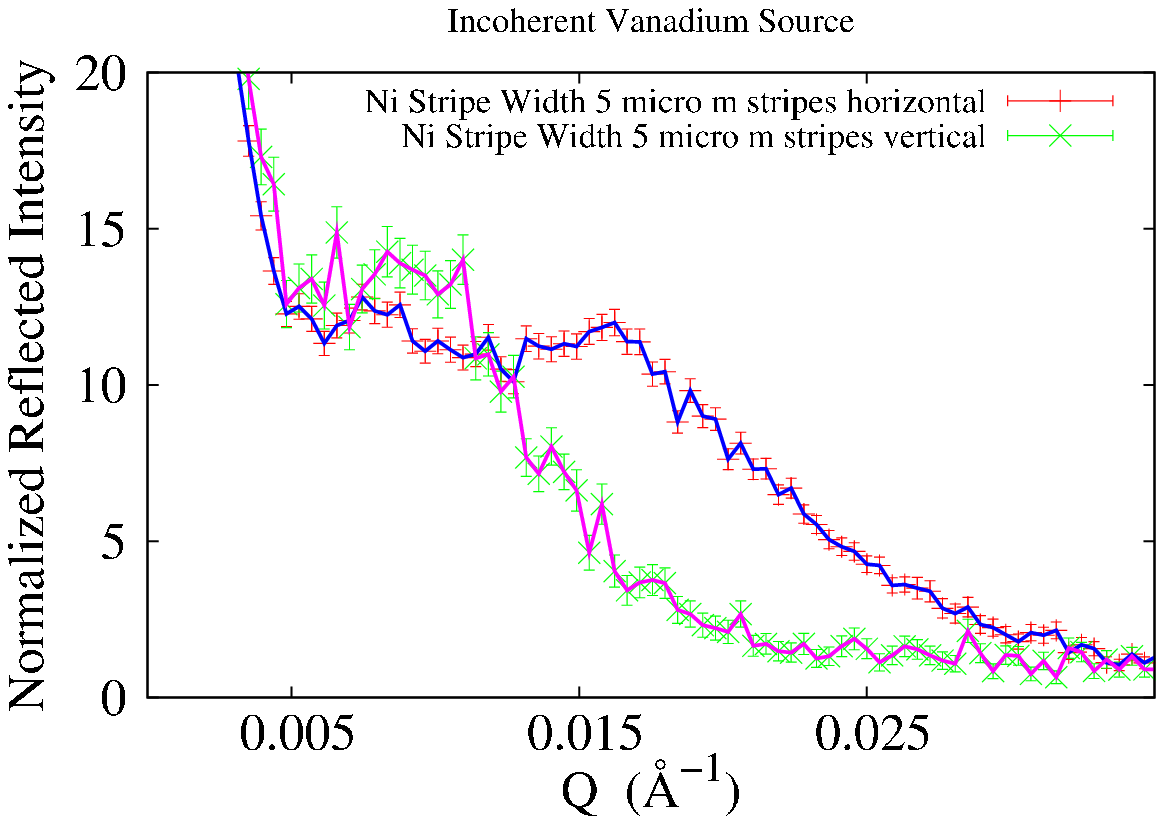}
\caption{\label{fig:17} 
 (Color online)  Specular reflection curves obtained with a polycrystalline cylinder of vanadium, a nearly perfect spin-incoherent, isotropic scatterer, instead of a monochromating crystal.  As discussed in the text, the results are again similar to those obtained with the PG monochromator despite the enormous difference in the instrumental incoherent resolution.  The rounding of the specular curve through characteristic critical angles, in comparison to corresponding curves obtained with the PG or Si monochromators, is due to the relaxed instrumental beam resolution -- but this does not obscure the clear difference between the case where the packets effectively average over the grating stripes and troughs and that in which they do not.  The overall shape of the specular reflectivity curve in the neighborhood of the critical angle is a direct ``yes/no'' qualitative indicator of coherence as has been emphasized throughout this section.  (Note that in this particular set of measurements the 5 + 5 = 10 micron grating was used.)}
\end{figure}

The standard focusing PG monochromator in the guide cut for the MAGIK reflectometer on the new NG-D guide at the NCNR was replaced with a dual component device consisting of a smaller monolithic piece of curved PG adjacent to a 1.25 inch diameter by 6 inch high cylinder of pure vanadium (one or the other component could be translated into the guide beam at a time).  The PG was used to align the gratings for subsequent experiments using the vanadium cylinder as the sole source of neutrons.  The Be filter, normally in a position between the monochromator and first of the pair of slits just upstream of the sample was removed.  Instead, a Be filter (at room temperature) was placed after the grating sample.  The polycrystalline Be filter effectively removes, by elastic Bragg reflection, neutrons of wavelength less than about 4 $\AA$ wavelength.  In addition, following the Be filter was placed a Fe/Si supermirror (3 x Ni critical angle) to deflect out all neutrons with nominal wavelengths greater than about 6 $\AA$ (the supermirror was set at a critical $Q$ value approximately that of ordinary Ni where both spin states are reflected).  To summarize, the instrumental configuration was: vanadium $\rightarrow$ pair of slits $\rightarrow$ grating sample $\rightarrow$ single slit $\rightarrow$ Be filter $\rightarrow$ supermirror deflector $\rightarrow$ single slit $\rightarrow$ ${}^3$He detector.  Except for the pair of slits, there was no other optical device between the vanadium source and grating sample.  This was an extremely simple set-up and the one most likely to prepare relatively undistorted wave packets with spherical wavefronts each having a radius of approximately 2 meters incident on the grating.  The post selection of wavelength range by the combination of Be filter and supermirror deflector resulted in a fractional wavelength bandwidth of about 0.4, from 4 to 6 $\AA$ centered about 5 $\AA$.  Although the intensity incident on the grating was lower by many orders of magnitude compared to that from a PG monochromating crystal, a usable number of neutrons -- approximately 1.4 counts/second -- was still obtained with an acceptable signal-to-noise ratio for specular reflectivity measurements in the region of the critical angles of interest.  It is of interest to note that at an incident intensity of 1.4 counts/sec, successive neutrons with  nominal wavelengths of 5 $\AA$ were separated on average by 565 meters!  (It is especially difficult to imagine any overlap of neutron wave packets in this case where one neutron has traversed the approximately 3 meter total distance between V source and capture in the detector -- via a nuclear reaction involving one neutron, one nucleus, and a specific number of reaction products -- long before another enters the instrument!)  Results obtained with the V source are summarized in Figure \ref{fig:17}.

\subsection{Comparison of the incoherent instrumental beam $Q$ resolution to the transverse coherence of the individual neutron wave packet }
Table \ref{table:2} compares the incoherent instrumental beam resolution for various combinations of apertures (defining the angular divergence of the beam) and monochromators or other source (predominantly determining the wavelength resolution).  ($\Delta Q$ was calculated using Equation \ref{eq:12} for a value of $Q = 0.015$ inverse Angstroms where the wavelength $\lambda = 5\AA$ and the corresponding value of $\theta$ is then 0.006 radians.) Values for the transverse coherence length which \textit{would} be predicted if equated to $1 /  \Delta Q$ -- as derived from the incoherent instrumental beam resolution -- are significantly smaller than the effective transverse coherent length observed in the measurements based on the specular reflectivity as reported in this paper (approximately 1 micron).  Despite the enormous difference (up to a factor of nearly 10) in the instrumental incoherent resolution of the beam for the various reflectometer configurations represented in the Table, the effective transverse coherence length obtained from the specular reflectivity measurements is essentially the same (see, again, Figures \ref{fig:13} and \ref{fig:16}: the data for the vanadium source shown in Figure \ref{fig:17} is consistent as well except that a factor of two adjustment is necessary in comparison because, in this one particular measurement, the 5 + 5 = 10 micron grating was used instead of the 10 + 10 = 20 micron grating employed in all of the other cases).  Recall from the analysis of the results presented in Figure \ref{fig:12} above that changing the grating period by a factor of two (from 20 to 40 microns) is all that is necessary to create an incoherent condition (in the case where the incoherent instrumental beam resolution for the data in Figure \ref{fig:12} is the same as the tightest value appearing in Table \ref{table:2} -- W = 0.2 mm).   Although for the very tightest slit widths $1 / \Delta Q$ is close to the value deduced for $\Delta r_\perp$ from the specular reflection measurements described above, coherent averaging in specular reflection continues to be observed despite increasing $\Delta Q$.  That is to say, if $\Delta r_\perp$ were proportional to $1 / \Delta Q$, then changing $\Delta Q$ to be a factor of two larger than that which was used in the rotation measurements for the 20 micron period grating (represented in Figure \ref{fig:12}) should suffice to ensure a loss of coherent averaging in the specular reflection process -- but this is not, evidently, observed.

\begin{table}
\caption{\label{table:2}
Comparison of incoherent instrumental beam resolution for various combinations of apertures (defining the angular divergence of the beam) and monochromators or other source (predominantly determining the wavelength resolution).  $\Delta Q$ was calculated using Equation \ref{eq:12} for a value of $Q = 0.015$ inverse Angstroms.  The wavelength $\lambda$ is 5 \AA.  The corresponding value of  
$\theta$ is then 0.006 radians.  Values for the transverse coherence length which \textit{would} be predicted if equated to $1 / \Delta Q$ -- as derived from the incoherent instrumental beam resolution -- are significantly smaller than the effective transverse coherent length observed in the measurements based on the specular reflectivity as reported in this paper -- approximately 1 micron.  Despite the enormous difference in the instrumental incoherent resolution of the beam for various reflectometer configurations, the effective transverse coherence length obtained from the specular reflectivity measurements is essentially the same (as described more thoroughly in the text).}
\begin{ruledtabular}
\begin{tabular}{lllll}
  $\Delta  \lambda / \lambda$ &        $W$ &       $\Delta \theta$ &            $\Delta Q$ &                      $1 / \Delta Q$ \\       
                           {} &  (mm)      & (mrad) &      $(\AA^{-1})$ &                    $ (\mu m)$ \\
  \hline
              0.01 (PG) &   0.2 &      0.13 &    3.58 x $10^{-4}$  &               0.280 \\
              0.01 (PG) &   0.4 &     0.26 &   6.67 x $10^{-4}$    &             0.150 \\
              0.01 (PG) &   1.0 &     0.71 &   17.8 x $10^{-4}$    &             0.056 \\
              $< 0.01$ (Si) &  0.5 &      0.33 &   8.46 x $10^{-4}$  &              0.118 \\
              0.4 (V)  &       1.0 &     0.71 &   62.6 x $10^{-4}$ &                0.016 \\
\end{tabular}
\end{ruledtabular}
\end{table}

\section{\label{sec:nonspecular_analysis}Analysis of non-specular reflectivity}
In addition to the specular reflectivity, it is possible to observe well-defined diffraction peaks at discrete values of $Q_x = 2\pi/D_x$ .  Practically, it is advantageous to access these reflections at glancing angles in the reflection geometry at a value of $Q = Q_z + 2\pi/D_x$.  Conventionally, the observed or measured width of such a reflection is presumed to be a convolution of a ``natural'' width related to the degree of correlation over the grating periods and the incoherent instrumental resolution (IIR) at a given location in reciprocal space.  However, in the present work we are assuming the grating to be perfectly periodic (see earlier section on the preparation and characterization of the gratings) and will attribute the natural linewidth soley to the intrinsic transverse coherence length of the neutron wave packet.

Besides the analysis of the linewidth of the grating reflections performed in the work by Salditt et al., cited earlier \cite{salditt1994x}, other groups have studied grating structures with x-rays and/or neutrons \cite{munter1994reflection,richardson1997study,tolan1992x,tolan1994x}.  In the work of Tolan et al., \cite{tolan1992x,tolan1994x}, the grating truncation rods were investigated and quantitative theories developed to account for both specular and off-specular scattering from the laterally structured surface.  In our measurements of non-specular line widths, we have avoided orientations with the grating stripes tilted off the perpendicular to the scattering plane defined by the nominal $\mathbf k_I$ and $\mathbf k_F$.  Thus, neither do we observe nor deal with the possible intersection of the vertically elongated IIR ellipsoid and the truncation rods which run parallel to the specular ridge (at reciprocals of the grating period along the pertinent in-plane direction).

The $x$-component of the wavevector transfer is given by (the effect of the out-of-scattering plane angle alpha appearing in Equation \ref{eq:4} is negligible in this analysis and is therefore omitted)
\begin{equation}
  \label{eq:13}
  Q_x = k_{Fx} - k_{Ix} \approx k(\cos \theta_F - \cos \theta_I)
\end{equation}
 
For the present case, the wavevector magnitude is practically independent of the angles $\theta_I$ and $\theta_F$ so that the corresponding uncertainty for the incoherent instrumental resolution
\begin{widetext}
\begin{equation}
  \label{eq:14}
  (\Delta Q_x)^2 = \left(\frac{\partial Q_x}{\partial k}\right)^2(\Delta k)^2 +
    \left(\frac{\partial Q_x}{\partial \theta_I}\right)^2(\Delta \theta_I)^2 +
    \left(\frac{\partial Q_x}{\partial \theta_F}\right)^2(\Delta \theta_F)^2 + 
    2C\frac{\partial Q_x}{\partial \theta_I}\frac{\partial Q_x}{\partial \theta_F}\Delta \theta_I \Delta \theta_F
\end{equation}

is approximately given by
\begin{equation}
  \label{eq:15}
   (\Delta Q_x)^2 \approx Q_x^2(\Delta k/k)^2 + k^2 \sin^2 \theta_I (\Delta \theta_I)^2 + k^2 \sin^2 \theta_F (\Delta\theta_F)^2 + 2Ck^2 \sin \theta_I (\Delta\theta_I) \sin \theta_F (\Delta\theta_F)
\end{equation}

\end{widetext}
(If $\theta_I$ and $\theta_F$ are completely correlated, then the coefficient $C$ is $\pm 1$, whereas if completely uncorrelated, $C$ would be zero.)

As discussed previously in another section, for the slit geometry employed in both the specular and non-specular measurements described here, the angular distribution of neutron trajectories in the incident beam can be accurately described with ray optics.  It is possible to calculate, geometrically, a value of the FWHM of an angular detector scan through the incident beam (no sample) that agrees with the measured value to within a few percent.  On the other hand, it can be difficult to accurately predict the IIR contribution to the measured angular width in a rotational scan of the sample at a fixed detector position or scattering angle.  This is primarily due to possible distortions of the sample substrates from perfect flatness and footprint corrections required for finite sample size at sufficiently low glancing angles.  Furthermore, for grating periods in the intermediate range between the incoherent and coherent limits, as observed in the specular measurements described above, the shapes and positions of the non-specular diffraction lines can be significantly affected.

To avoid the complications described above, non-specular scattering from the 10 + 10 = 20 micron period grating, for which the shapes and positions of the grating diffraction lines were not significantly distorted, was analyzed according to the following procedure.

The relationship among the measured (MS), incoherent instrumental resolution (IIR) (for the x-component of $Q$), and natural (NAT) components of the grating line widths is given by
\begin{equation}
  \label{eq:16}
  (\Delta \theta_\mathrm{MS})^2 = (\Delta \theta_\mathrm{NAT})^2 + (\Delta \theta_\mathrm{IIR})^2
\end{equation}
where by taking these quantities to be related in quadrature we are assuming the natural grating line width and incoherent instrumental resolution to be completely uncorrelated with each other.  We demonstrated earlier that in the case of a perfectly flat substrate we can calculate the IIR term from known slit widths and beam path lengths to an accuracy of the order of 1 percent (see Eqs. \ref{eq:6}-\ref{eq:9} and accompanying discussion).  Thus, the uncertainty $\delta (\Delta \theta_\mathrm{IIR})$ is negligible for a flat substrate.  If we knew a priori what the substrate distortion was, we could also account for it, in our geometrical ray analysis, by adding another term on the RHS of Eq. 6 which would represent an effective broadening or focusing angle corresponding to an overall convex or concave curvature of the surface (positive or negative value), respectively.  This assumes that the distortion is relatively smooth so as not to introduce any significant nonuniform structure in the shape of the rocking curve.  (For the substrates used in the measurements described herein, no such irregular rocking curves were observed.)  We can write this additional substrate distortion term explicitly in the expression for the overall $\Delta \theta_\mathrm{IIR}$ as

\begin{equation}
  \label{eq:17}
  (\Delta \theta_\mathrm{IIR})^2 = \left(\Delta \theta_\mathrm{IIR}{}^\text{FLAT SUB} + \Delta \theta_\mathrm{IIR}{}^\text{DISTORTED SUB}\right)^2
\end{equation}

If the distortion term is not known beforehand, we can, alternatively, solve for it by performing a set of diffraction linewidth measurements, for a proper selection of different slit settings, as we will now describe. 

As discussed previously, we associate the natural linewidth with the transverse coherence of the neutron wave packet and not the grating stripe correlations (which have been shown to be of sufficiently long range in a previous section).  If measurements of the widths are performed at two different settings of the instrumental resolution, determined primarily by the angular divergences of the beam, we can solve a simultaneous pair of algebraic equations for the common natural line width term.  We can do so for two grating reflections, the m = - 1 and the m = + 1 on either side of the specular reflection (m = 0).  The two instrumental collimations chosen were: a) 0.0074 and 0.0094 degrees; and b) 0.0074 and 0.02820 degrees --  upstream and downstream of the sample, respectively, in both cases.  For the 10 + 10 =20 micron Ni stripe grating chosen for this measurement, the first order grating reflections are centered about an incident beam angle of 0.3 and 0.5 degrees on either side of the specular position of 0.4 degrees (SA = 0.8 degrees).  Note that by selecting two different instrumental angular resolutions such that the upstream slits remained constant while only the downstream angular divergence varied, it was ensured that the footprint of the beam on the sample substrate plane remained constant for a given grating reflection.  However, the footprints for the m = -1 and m = +1 reflections, once again centered about the incident beam angles of 0.3 and 0.5 degrees, respectively, differed significantly -- approximately 0.73 of the length along the 7.5 cm diameter of the circular Si substrate was illuminated at 0.3 degrees for the m = -1 reflection whereas only 0.44 of that length is illuminated at 0.5 degrees for the m = +1 reflection.  Thus, the possibility that the angular term attributed to substrate distortion must be treated as being potentially different for the two cases.  Consequently, we will analyze the two grating reflections independently of one another.

Fortunately, it is still possible to solve for the common natural width as well as the substrate distortion term separately for each of the two grating reflections -- provided we use a calculated value for the instrumental angular divergence contribution defined by the aperture widths corresponding to a perfectly flat substrate.  We demonstrated earlier that it was possible to calculate this contribution with good accuracy (uncertainties of the order of a percent).  Table \ref{table:3} summarizes the measured widths from which the natural widths $(\Delta \theta_\text{NAT})$ and substrate distortions $(\Delta \theta_
\text{IIR}{}^\text{DISTORTED SUB})$ are calculated via Equations \ref{eq:16} and \ref{eq:17} (as well as \ref{eq:6} through \ref{eq:9}).  The incoherent instrumental resolution widths for a rocking curve scan of a perfectly flat sample are calculated to be 0.01185 degrees and 0.0234 degrees for collimations ``a'' and ``b'', respectively, referred to in the preceding paragraph.  In subsequent calculations where the widths are converted to angular measure from wavevector transfer units of reciprocal Angstroms along the $x$-direction, $Q_x$, the relationship among scattering, incident,and final angles is needed to apply Equation \ref{eq:13}.  This relationship is provided by
\begin{equation}
  \label{eq:18}
  \text{SA} = \theta_I + \theta_F
\end{equation}

\begin{table}
\caption{\label{table:3}
Measured widths of the nonspecular reflections along $Q_x$ for the 10 + 10 = 20 micron Ni stripe grating.  LHM and UHM indicate lower and upper half-maximum positions, respectively.  A zero offset correction of 0.002 deg has been applied uniformly to all peak positions.  The position of m = 0 is at 0.400 deg and the scattering angle is 0.800 deg.  The accuracy of the numbers in the table is of the order of 1 \%.  Also included in the table are the peak positions and widths for a flat substrate.}
\begin{ruledtabular}
\begin{tabular}{rcccccc}
 m  &   Peak &   Collimation & $\Delta \theta_\text{IIR}^\text{FLAT SUB}$ &       UHM  &-  LHM =&    $\Delta \theta_\text{MS}$ \\                         
 {} &         (deg)   &         FWHM  &         FWHM      &     (deg)  &   (deg) &      FWHM  \\
 {} &          {}     &        (deg)  &          (deg)    &      {}    &     {} &      (deg)  \\                                              
\hline
             -1  &      0.304   &     0.0074 / 0.0094 &     0.01185 &          0.313 &    0.294 &          0.019  \\      
             +1  &      0.498   &     0.0074 / 0.0094 &     0.01185 &          0.506 &    0.490 &          0.016   \\     
             -1  &      0.302   &     0.0074 / 0.0282 &     0.0234  &           0.317 &     0.287 &           0.030  \\      
             +1  &      0.497   &     0.0074 / 0.0282 &     0.0234  &           0.508 &    0.486 &           0.022  \\
\end{tabular}
\end{ruledtabular}
\end{table}

\begin{figure}
\includegraphics[width=\linewidth]{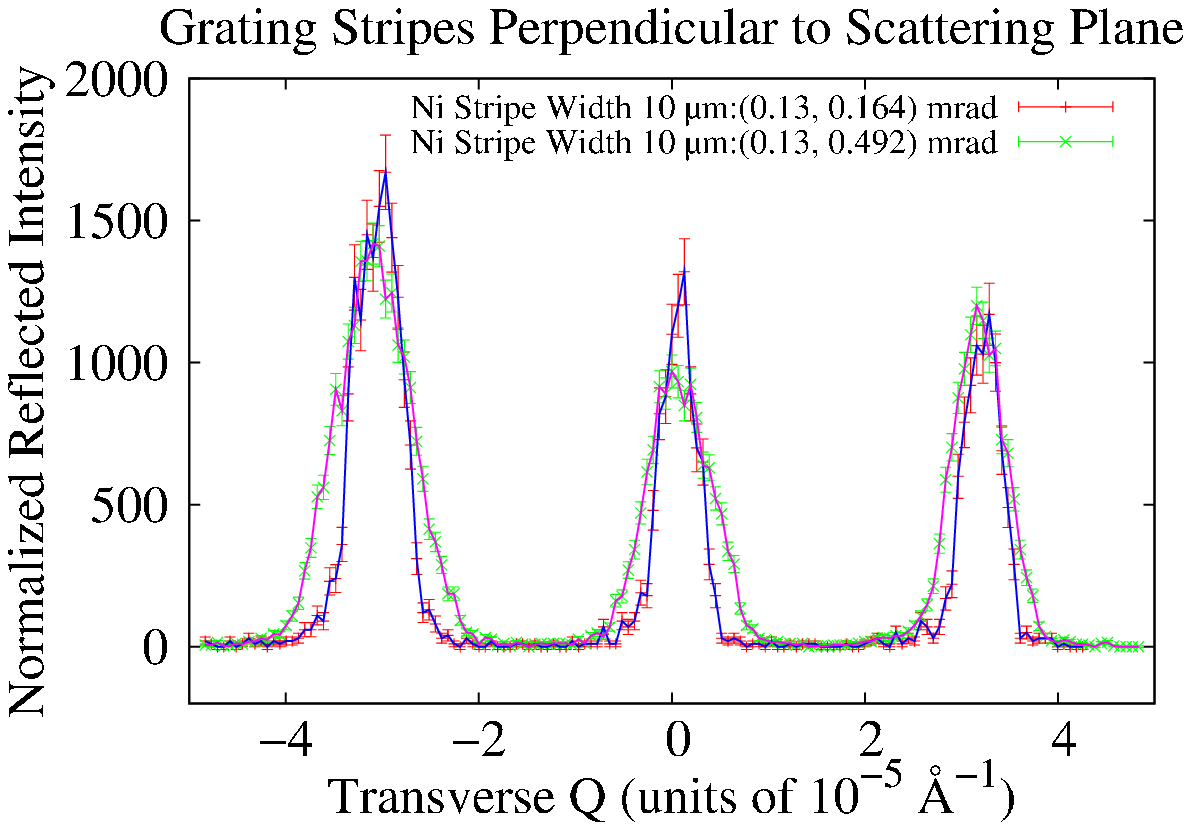}
\caption{\label{fig:18} 
(Color online)  Non-specular neutron diffraction grating lines, m = +/- 1, about the specular ridge line, m = 0.  The two sets of data correspond to different incoherent instrumental resolution values as discussed in the text.}
\end{figure}

In Figure \ref{fig:18} are shown the two measured scans along $Q_x$ at these two different beam resolutions.   Two diffraction orders, m = +/- 1, are observed on either side of the specular ridge which is labeled m = 0.

Table \ref{table:4} gives the natural widths along with the instrumental width associated with substrate distortion as extracted from the measured widths for both the m = -1 and m= +1 grating reflections.  The widths are given in angular measure as well as in reciprocal Angstroms for $Q_x$.  The transverse coherence length of the neutron wave packet and its projection on to the plane of the sample surface were calculated from the following equations (refer back to Figure 7):
\begin{equation}
  \label{eq:19}
  \Delta Q_\text{NAT} = 2\pi / L_x
\end{equation}

and
\begin{equation}
  \label{eq:20}
  \Delta r_\perp = L_x \sin \theta_I
\end{equation}

\begin{table}
\caption{\label{table:4}
 Natural widths corresponding to neutron transverse coherence lengths as extracted from a measurement of the widths of the m = -1 and m = + 1 nonspecular reflections for the 10 + 10 = 20 micrometer Ni stripe grating -- according to the simplistic, approximate theory described in the text.  The negative sign indicated for the width associated with substrate deformation in the case of the + 1 reflection is attributed to focusing by a net concave, as opposed to convex, curvature.  (The flatness tolerance that is specified to the manufacturer of the Si substrates is +/- 0.01 degree deviation of any local normal from that of the mean surface normal.)  The accuracy of the numbers representing angular measurements in the tables is of the order of 1 \%.  As it turns out, the values for the neutron transverse coherence length predicted by Equations \ref{eq:19} and \ref{eq:20} which appear in this Table agree remarkably well with the results of the more rigorous analysis given below and in the accompanying paper, Part II.  From Table \ref{table:4}, $\Delta r_\perp  = 1.34 \pm 0.2$ microns, on average.}
\begin{ruledtabular}
\begin{tabular}{rlllll}
m  &  $\Delta \theta_\text{IIR}^\text{DISTORTED SUB}$ &  $\Delta \theta_\text{NAT}$ & $\Delta Q_\text{NAT}$ & $L$ & $\Delta r_\perp$ \\
{} &             (deg)              &      FWHM   &      FWHM   &      ($\mu\text{m}$) &     ($\mu\text{m}$) \\
{} & {} &                                 (deg)   &       $(\AA^{-1})$ & {} & {} \\
\hline
           -1 &      0.0063 &              0.0067 &        $2.16 \times 10^{-6}$ &       291. &        1.54 \\
           +1 &     -0.0072 &              0.0150 &        $4.83 \times 10^{-6}$ &       130. &        1.13 \\
\end{tabular}
\end{ruledtabular}
\end{table}

The relationship between the natural reflection width and projected coherence length $L$ in Equation \ref{eq:19} is derived from the treatment of diffraction for a one-dimensional, periodic rectangular potential within the Born approximation assuming an undistorted incident plane wave.  Equation \ref{eq:20} follows from the schematic representation of Figure \ref{fig:7}.  Once again, these simplistic models reflect neither the neutron's wavepacket character nor the significant distortion which occurs in the neighborhood of the material grating during the diffraction process.  A more realistic treatment is given in a following section and the accompanying theory paper, Part II.  Nonetheless, as it turns out, the values for the neutron transverse coherence length predicted by Equations \ref{eq:19} and \ref{eq:20} which appear in Table \ref{table:4} agree remarkably well with the results of the more rigorous analysis given below.  From Table \ref{table:4},  $\Delta r_\perp  = 1.34 \pm 0.2~ \mu\text{m}$, on average.  (Note that, thus far, no consideration has been given to the possibility that a distribution of different wave packet sizes exists within the collection of neutrons comprising the beam -- due, for example, to reflection from a mosaic monochromator crystal made up of a distribution of different microcrystallite sizes.)  

Table \ref{table:5} gives the widths for the m = -1 and m = + 1 nonspecular reflections of a 10 + 10 = 20 micrometer Ni stripe grating as predicted by the more rigorous wave packet theory presented in the section after next below and, in more detail, in the associated paper, Part II -- in this case an average value of approximately 1.7 +/- 0.3 microns for the transverse coherence length of the neutron wave packet is obtained.

\begin{table}
\caption{\label{table:5}
Widths predicted for the m = -1 and m = + 1 nonspecular reflections of a 10 + 10 = 20 micrometer Ni stripe grating, according to the more rigorous wave packet theory presented in the text (see section after next below) and, in more detail, in the associated paper, Part II, given the values of the neutron wave packet transverse coherence length listed in the Table.  This represents an average value of approximately 1.7 +/- 0.3 microns for the transverse coherence length of the neutron wave packet.
}
\begin{ruledtabular}
\begin{tabular}{rll}
m  & $\Delta \theta_\text{NAT}$ & $\Delta r_\perp$ \\
{} &             FWHM           &  $(\mu\text{m})$ \\
{} &             (deg)          &        {}        \\
\hline
           -1  &       0.007 &                   1.41 \\
           +1  &      0.021  &                  1.41 \\
           -1  &       0.005 &                   1.99 \\
           +1  &      0.0152 &                 1.99 \\
\end{tabular}
\end{ruledtabular}
\end{table}

\section{\label{sec:forming_wavepacket}Forming of the wave packet shape by various neutron optical elements of the instrument}
As presented in the preceding sections, our measurements give a value of approximately one micron for the effective neutron transverse coherence in specular (as well as nonspecular) reflection from gratings on a nominally flat substrate.  Can we identify what elements of the reflectometer have a significant role in forming or limiting this length?  After all, the underlying premise is that the coherence of the wave packet is not an intrinsic (or possessed) property of the neutron but one that is instead imparted by the optical elements of the instrument with which the neutron interacts.  

Having already discounted the slit widths as being too large to significantly affect the transverse coherence of the neutron wave packet in the present case, we consider the influence of other neutron optical elements in the reflectometer.  From analysis by Fresnel-Huygens construction, it is expected that a neutron wave packet reflected from a perfect single microcrystallite of PG would emerge with a collection of wavefronts that are laterally coherent over dimensions of the order of the size of the particular crystallite.  The PG microcrystallites typically have dimensions between one and ten microns or so \cite{simonis2002stm}.  However, as shown earlier, similar values of the effective transverse coherence length were obtained using a Si perfect single crystal (of macroscopic dimensions of several centimeters) and an incoherent scattering vanadium source in place of the PG.  For the vanadium source, an emanating spherical wavefront should be coherent over a distance of the order of 100 microns at a distance of 2 meters from the grating (as discussed below).  These results imply the existence of another limiting factor on the effective transverse coherence.

It might be also asked why, for example, in single slit diffraction experiments with neutrons, as carried out by Zeilinger et al. \cite{zeilinger1988single}, slit widths up to 90 microns produced a diffraction pattern almost indistinguishable from the ideal theoretical prediction, thus implying a transverse neutron coherence length of at least comparable value.  The preparation of the neutrons in the instrument used to perform these slit diffraction measurements did not differ so much from that in the present reflection experiments -- however, the diffraction from a slit with no material between the sharp, parallel absorbing edges defining an aperture is significantly different than the process of specular reflection involving a continuum of scattering material across an entire contributing surface area. 

One possibility might then be the deviation from perfect flatness of the neutron guide tubes -- as well as the Si substrates on which the Ni grating structures were deposited.  It is reasonable to expect that the maximum degree of constructive interference for a specular reflection process would occur when the wavefront of the radiation and the continuous material surface are as conformal with respect to one another as possible.  Of course, other, smaller length scales corresponding, for example, to roughness of the material surface on a nanometer or atomic scale, may be present simultaneously -- but on this level, insofar as specular scattering is concerned, the SLD is assumed to be effectively averaged over.  It is the longer length scale associated with variations in the mean normal of either the wavefront or material surface that is pertinent to the present discussion. 

\begin{figure}
\includegraphics[width=\linewidth]{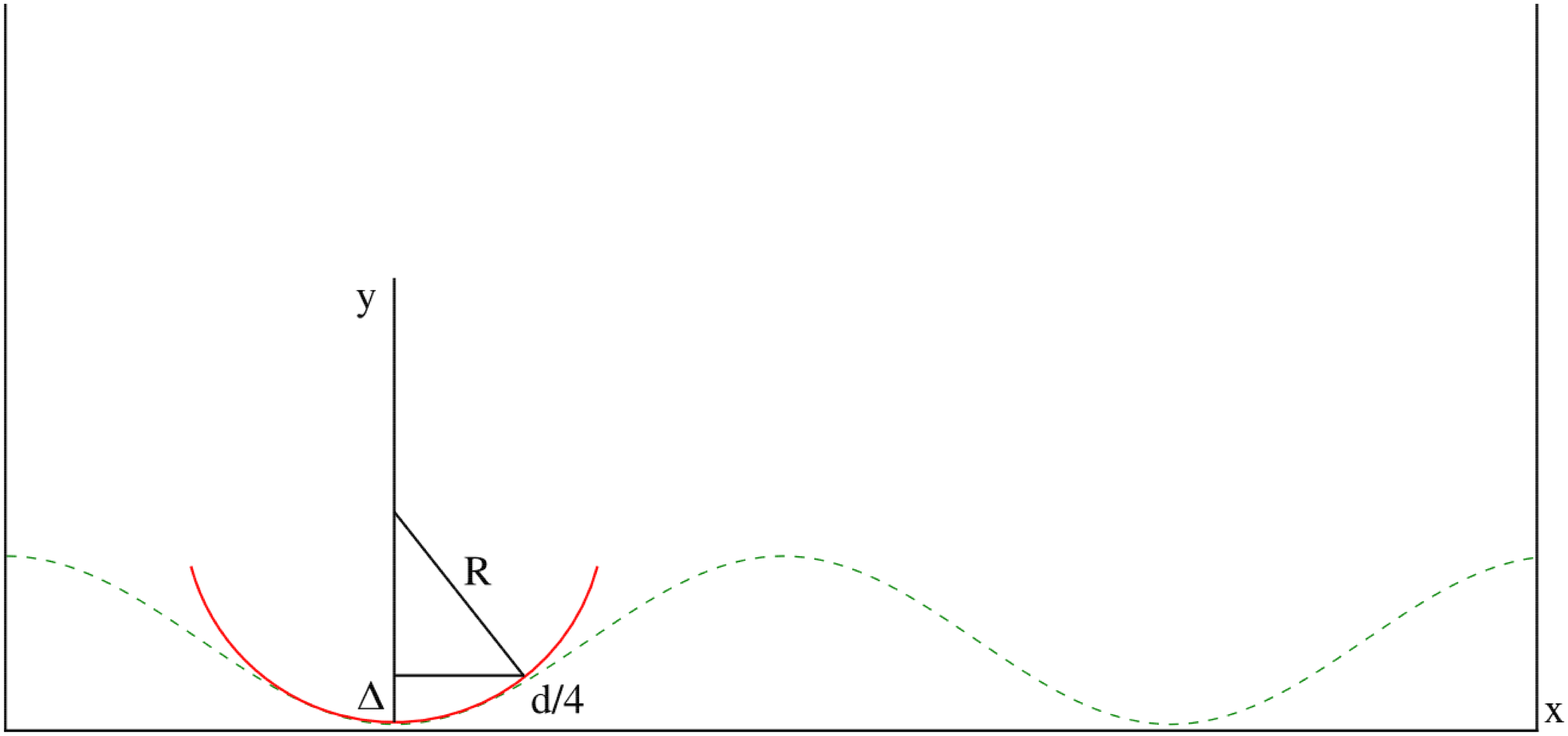}
\caption{\label{fig:19} 
 (Color online)  Schematic representation of two primary length scales, $X$ and $d$, corresponding to the chord of a circular wavefront of incident radiation and the period of a sinusoidal material surface undulation, respectively.  The horizontal $x$- and vertical $y$-axes represent distance in two orthogonal directions but are not on the same scale -- thus, the circular radiation wavefront and material surface boundary are significantly distorted from realistic values in the diagram.  These particular measures of length might be considered in attempting to understand, qualitatively at least, the role which the material surface might play in determining the \textit{effective} transverse coherence length $r_E$ of a wavefront for the process of specular reflection.  See the text for a more extensive discussion.}
\end{figure}

To examine such a possibility, consider Figure \ref{fig:19} which is a schematic representation of two primary length scales which might be involved in determining the {\textit{effective}} transverse coherence length $r_E$ of a wavefront (associated with one possible component within the composite wave packet).  
One length is associated with the free neutron wave packet itself which we designate $r_\text{WP}$ and represent here as the chord of a circular wavefront of radius $R$ -- one half of this cord is shown in Figure \ref{fig:19} with a projected length of $d/4$ along the $x$-axis.  The other length is taken to correspond to an undulation of a material surface from which the neutron wave is specularly reflected -- for simplicity, the material surface boundary is here represented in Figure \ref{fig:19} as a sine curve with period $d$ where the greatest deviation angle $\alpha$ of the local surface normal from that of the mean occurs at $d/4$.  The equation of the line representing the surface along one direction and its slope are given by
\begin{equation}
  \label{eq:21}
  y = y_0 + y_0 \sin [(2\pi x/ d) - (\pi/2)]
\end{equation}

and
\begin{equation}
  \label{eq:22}
  dy/dx = y_0 \cos [(2\pi x/d) - (\pi/2)](2\pi/d) = \tan \alpha
\end{equation}
respectively.  At $x = d/4$, $\tan \alpha = 2\pi/d$.  The half chord length $X$ of the circular wavefront, on the other hand, is given by
\begin{equation}
  \label{eq:23}
  X = [2R\Delta - \Delta^2]^{1/2} 
\end{equation}

As a rough, qualitative measure of where the wavefront and surface get out of phase with one another and the condition for coherent specular reflection breaks down, suppose that the two curves intersect at $\pm d/4$ and let $\Delta = y_0 = \lambda$, the neutron mean or nominal wavelength.  For the vanadium source measurements described above, $R$ is approximately 2 meters and the mean neutron wavelength is 5 $\AA$ so that Equation (\ref{eq:23}) gives $2X$ equal to about 90 microns.  On the other hand, the deviation angle $\alpha$ is typically 0.01 degree for the silicon wafers used as substrates for the grating structure support (and also, incidentally, for the glass substrates of the neutron guide system) so that given a surface peak-to-valley amplitude of the order of 2 $\AA$, for example, Equation (\ref{eq:21}) predicts $d/4$ to be only about one micron.  Therefore, in this simplistic model at least, it is the surface curvature which limits the effective transverse coherence in the interaction of the wavefront of the packet with the wavy material surface.

A model-independent way for incorporating these two independent length scales into the coherent extent of the wave/surface boundary is to use their harmonic mean.  Thus we define an effective coherence length $r_E$ by
\begin{equation}
  \label{eq:24}
  1 / r_E = 1 / r_\text{WP} + 1 / r_s 
\end{equation}

or
\begin{equation}
  \label{eq:25}
 r_E  = r_\text{WP} r_s / ( r_\text{WP} + r_s )
\end{equation}
where the latter length, $r_s$, is the length associated with the material surface.  If $r_s \gg r_\text{WP}$ , then $r_\text{WP}$  is the limiting length whereas if  $r_s \ll r_\text{WP}$ , then $r_s$ limits the coherence in the specular reflection process.

We emphasize that the consideration of the degree of conformality between neutron wavefront and reflecting surface curvature, discussed above, as a possible limiting factor on the neutron transverse coherence is not needed to produce the value we report but is merely consistent with it.  That is, we have obtained through direct measurement a value of approximately one micron whatever processes or objects have in fact shaped the wave packet. 

\section{\label{sec:theory_perfect_grating}Theory of the specular scattering of a neutron wave packet by a perfect grating}
Previous work has described x-ray and neutron scattering from grating structures according to dynamical theory, but using plane wave functions to describe the incident radiation \cite{tolan1995x,ashkar2010dynamical}.  A more comprehensive treatment of the theory briefly summarized here in this section is given in the accompanying theory paper, Part II. 

We need a general approach to analyzing measured reflectivities for those situations where the incident wave (for a single incident quantum of radiation) is a wave packet. A full derivation of such a treatment is presented in Part II of this report [N.F. Berk], along with implementation details for its application to gratings. Here we only outline the concept and show formulas we have used for the present problem. The basic idea is simple enough: while the incident wave packet we wish to measure is not a stationary solution of the Schroedinger equation, the detection system and associated methods of analysis effectively treat the scattered wave as if it were, thereby, gating (or filtering) the wave packet   ̶ i.e., allowing only those plane wave constituents of the scattered wave packet consistent with the elastic condition   ̶ to be observed. The resulting gated scattering amplitudes (here, the reflection coefficients) thus are superpositions of constituent plane wave amplitudes with weights determined by the ``shape'' of the incident wave packet but restricted by the energy gating.

As discussed earlier in the Introduction, for the purposes here we need only consider the special case of reflection from a ruled grating with its stripes perpendicular to the slits defining the incident beam, effectively reducing the problem to a one dimensional grating of period Tg  along the $x$-axis, with two reciprocal space dimensions, $\mathbf k = ( k_z , k_x )$, where $k_z = k_z (k_x) = (k^2 - k_x^2)^{1/2}$  for given $k = k(E)$. In the plane wave problem, reflections occur only at $k_{xm} = k_x  + G_m$ for incident $k_x = k \cos \theta$, where 
$G_m = 2 \pi m / T_g$  is the $m$-th reciprocal lattice vector for a grating of period $T_g$ . The observable reflection amplitudes for an incident wave packet then turn out to be given by
\begin{equation}
  \label{eq:26}
  \tilde r_k(k_{xm}) = \sum_n \tilde A_k (\phi_{n-m})
    r_{kn}(\theta-\phi_{n-m})
\end{equation}

In this equation,  $r_{kn}( \theta - \phi)$ is the plane wave reflection amplitude at $k_{xn}$ for glancing incident angle
$( \theta - \phi)$, and
\begin{equation}
  \label{eq:27}                                                   
      \tilde A_k(\phi) = |sin(\theta)| A_k(\phi) / (|sin(\theta-\phi)| A_k(0))
\end{equation}
where (for any $\phi$),
\begin{equation}
  \label{eq:28}
    A(\phi) = (\sqrt{\sigma_L \sigma_T}/2 \pi) e^{-(1/2)(k\sigma_L)^2 (cos\phi-1)^2} e^{-(1/2)(k\sigma_T)^2 sin^2\phi}
\end{equation}
represents the Fourier content of an incident wave packet that is modeled by a Gaussian ``cigar'' having longitudinal (L) and transverse (T) lengths scaled by the standard deviations $\sigma_L$ and $\sigma_T$ , respectively. Note that

\begin{equation}
  \label{eq:29}
  A(\phi) = (\sqrt{\sigma_L \sigma_T}/2 \pi) e^{-(1/2)(k\sigma_T)^2 \phi^2}
\end{equation}
for sufficiently small $\phi$. The incident wave packet can be viewed as a bundle of \textit{virtual} incident plane waves. The angle $\phi_{n-m}$ is the solution of
\begin{equation}
  \label{eq:30}
  k \cos(\theta - \phi_{n-m}) + G_n = k_{xm}
\end{equation}
for fixed incident $\theta$, which describes the ``aliasing'' of the $n$-th plane wave reflection for a virtual plane wave incident at $\theta - \phi_{n-m}$ to the observed $m$-th reflection.  In particular, on the specular ridge, corresponding to $m = 0$,	 
\begin{equation}
  \label{eq:31}
  \tilde r_k(k_{x}) = r_0(k_{x}) + \sum_{n \neq 0} \tilde A_k(\phi_n)r_n(\theta-\phi_n)
\end{equation}
for the perfectly smooth ``grating'', since then $r_n = 0$ for all $n \neq 0$. That is, the wave packet shape is invisible to a perfectly smooth film, a result that can be shown quite generally for the gated wave packet. Otherwise, the ``observed'' reflection amplitude on the specular ridge is the plane wave amplitude plus a wave packet--weighted sum over all aliased reflection amplitudes.

\begin{figure}
\includegraphics[width=\linewidth]{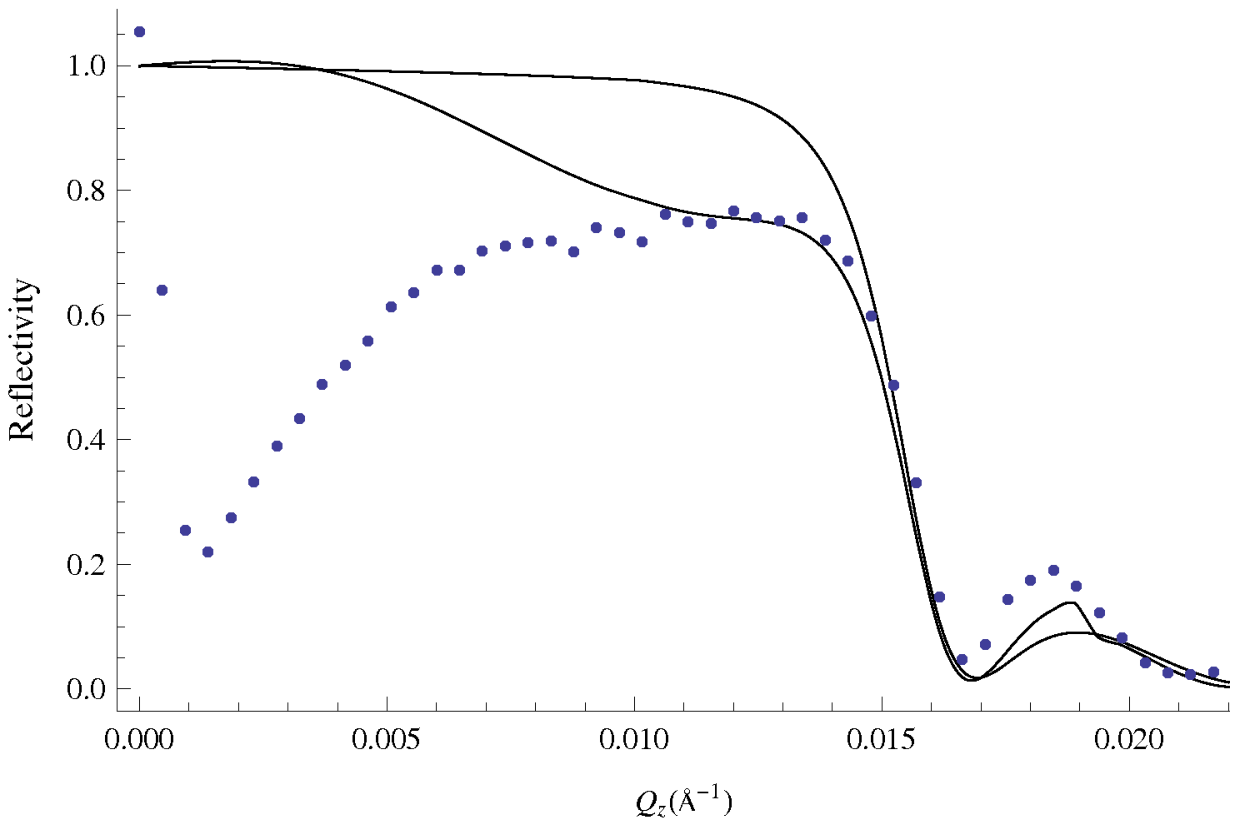}
\caption{\label{fig:20} 
(Color online)  Comparison of theory and experiment for specular reflection from the 10 + 10 = 20 micron period grating with Ni stripes perpendicular to incident nominal neutron wavevector.  The solid circular symbols represent the measured reflectivity (but without any beam footprint correction) whereas the solid line in close agreement with the data (approximately 75 \% reflectivity at the critical edge) is that calculated for an incident neutron represented by a Gaussian wave packet with a transverse coherence length of 1.4 microns (FWHM).  (The other solid curve is the specular reflectivity to be expected from a film of the same thickness but with the average SLD of the grating structure.)  Nonetheless, varying this value by the relatively small amount of plus or minus a micron does not have a significant effect on the specular curve in this range of Q about the critical angle.  On the other hand, the prediction of the same theory but for the non-specular, grating reflection linewidth is far more sensitive, and is in excellent agreement with measurement as presented in a previous section (see Table \ref{table:5}). }
\end{figure}

Figure \ref{fig:20} shows a comparison of theory and experiment for specular reflection from the 10 +10 = 20 micron period grating with Ni stripes perpendicular to incident nominal neutron wavevector.  
Good agreement between the wave packet theory described above and that measured is obtained for an incident neutron represented by a Gaussian wave packet with a transverse coherence length of 1.4 microns (FWHM).  Nonetheless, varying this value by the (relatively small) amount of plus or minus a micron does not have a significant effect on the specular curve in this $Q$-range about the critical angle.  On the other hand, the same wave packet theory developed here was also used to describe non-specular scattering, specifically to predict the natural line widths of the m = +/- 1 reflections for the 10 + 10 = 20 micrometer grating.  This theoretical prediction for the linewidth is considerably more sensitive and is in excellent agreement with the experimental results as shown previously in Table \ref{table:5}.

\section{\label{sec:conclusions}Conclusions}
By analyzing the specular reflectivity from a set of well-characterized diffraction gratings of different  periods, we have obtained a measure of the effective transverse coherence length of a neutron wave packet -- as prepared in a typical reflectometer configured for the study of the structure of condensed matter in thin film form.  The particular measure of transverse coherence obtained in this manner is indicative of the ability of the packet to effectively average over in-plane variations of scattering density in specular reflectivity measurements which are sensitive to the scattering density depth profile along the mean surface normal of the film structure.  It turns out that this measure of coherence, transverse or perpendicular to the mean wavevector of the packet, is of the order of one micrometer and is found to be largely independent of the incoherent instrumental beam resolution that describes the distribution of mean wavevectors of the individual neutron wave packets, the collection of which constitute the beam.  Although all of the optical devices of the reflectometer -- from liquid hydrogen moderator source, to guides, to monochromator crystal, to slits -- potentially play some role in simultaneously defining both the effective transverse coherence of each individual neutron wave packet as well as the angular divergence and wavelength spread of the beam made up of a collection of such wave packets, measurement of the individual packet and collective beam characteristics are to a degree separable under certain conditions, as we have described.  The effective neutron transverse coherence length obtained for the instrumental configuration described herein is very likely limited primarily by the waviness of the material substrate for the grating structure and/or that of the guide tubes.  Given that the transverse coherence of the wave packet is not an intrinsic or possessed property of the neutron but is, rather, shaped by the instrument, employing (for example) reflecting mirrors with flatter surfaces may achieve a greater effective coherence. 

Knowing some measure of the effective coherence of the packet, specifically its effective transverse coherence in specular reflection, in addition to the conventional incoherent instrumental beam resolution, a more accurate analysis of reflectivity data can be performed for material films where in-plane inhomogeneities of scattering length density exist.  In cases where the in-plane inhomogeneity of a film system is not too extreme and the length scales and compositional variation are sufficiently well-known, a wave packet description, such as discussed in the accompanying theory paper, Part II, may not be required to achieve a certain level of accuracy in extracting SLD profiles from specular reflectivity data.  In such circumstances it would suffice to perform an analysis based on the plane wave solution of the 1D time-independent Schroedinger equation, as conventionally done, but informed as to what extent the wave effectively averages across the surface (i.e., a single coherent average or incoherent sum of reflected intensities with proper area-weightings) -- while including a standard deconvolution of the incoherent instrumental beam resolution from the measured reflectivity.

In the Schroedinger formulation of quantum mechanics, the wavefunction is sometimes regarded more as a mathematical tool for predicting the outcomes of observations than as a physical entity existing in space \cite{styer2002nine}.  But scattering studies such as that performed here clearly demonstrate the efficacy in also regarding the wave packet as a measure of the extent of an effective volume of interaction between neutron and other material object in real space.  Further studies employing various grating structures with well-characterized properties may yield additional information about the composition and shape of neutron wave packets as quantum objects.

\section{Acknowledgments}
We acknowledge useful and informative discussions with P. Klosowski, P. Gehring, R. Cappelletti, R. Pynn, R. Dimeo, and R. Cubitt  The identification of any commercial product used in the course of the research reported herein does not in any way imply an endorsement thereof.  We are grateful to J. Copley for providing the vanadium source and acknowledge B.G. Jones for assistance in the fabrication of the gratings.  Some of the grating structures were prepared in part at the NIST Center for Nanoscale Science and Technology.

\appendix
\section{Details of the fabrication and characterization of the diffraction gratings}
Diffraction gratings were constructed of a variety of stripe materials deposited on 3 inch diameter by 1 mm thick Si substrates.  The stripes were nominally of rectangular cross section and 1000 Angstroms thick with the stripe width approximately equal to the spacing between stripes (so that the period of the grating is roughly twice the stripe width).  Stripe materials included both Ni and Permalloy, but most of the measurements reported on here are for the Ni stripe gratings.  Grating periods of 1600 (800 + 800), 800 (400 + 400), 400 (200 + 200), 200 (100 + 100), 100 (50 + 50), 50 (25 + 25), 20 (10 + 10), 10 (5 + 5) micrometers were fabricated.

\begin{figure}
\includegraphics[width=\linewidth]{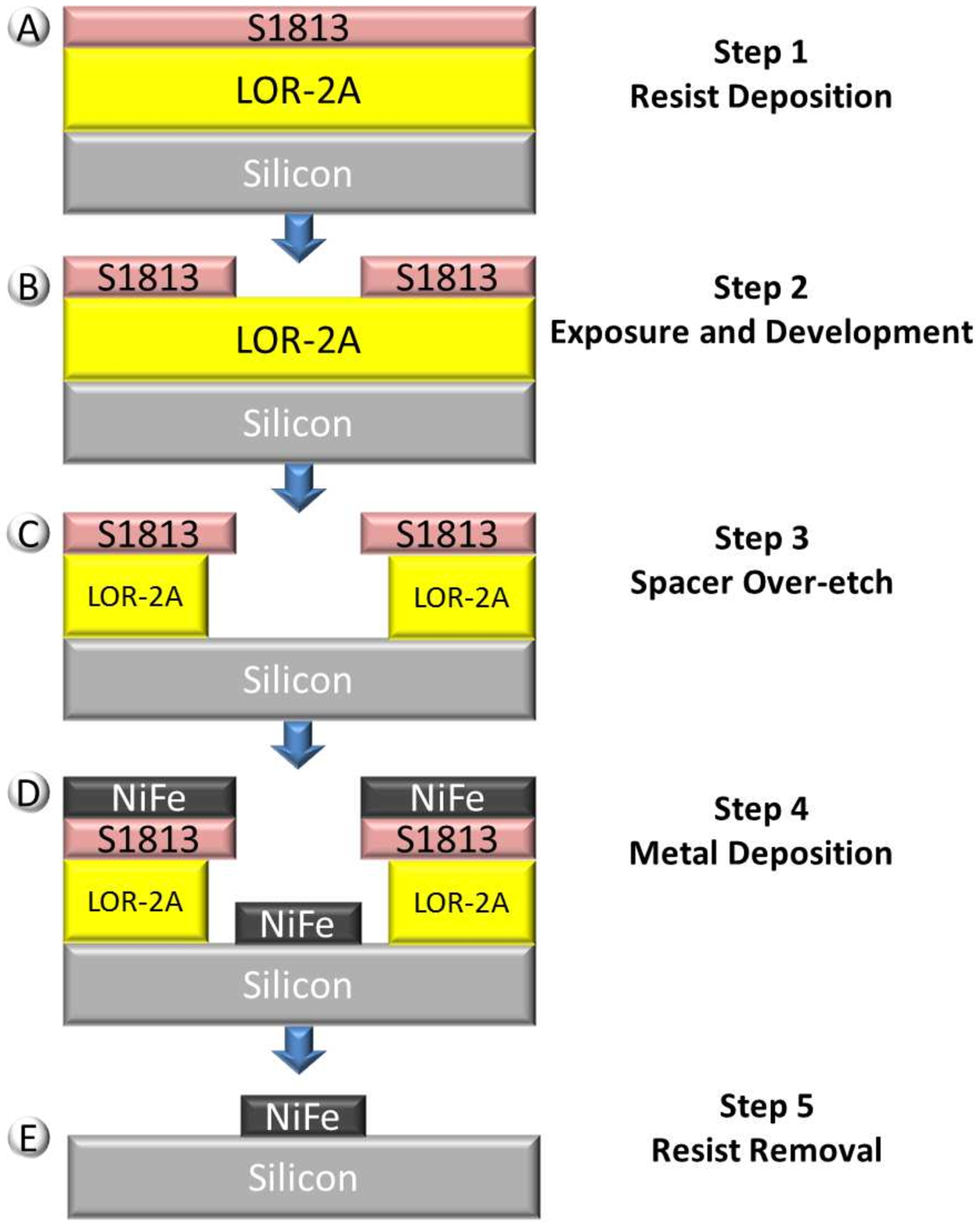}
\caption{\label{fig:A1} 
 A) A spacing layer followed by a resist layer are spin-coated. B) The resist is exposed using a mask and developed. C) The spacer layer is developed leaving a lip. D) Metal is deposited onto the sample (in most cases this was Ni -- however, some gratings were made with Permalloy, a NiFe alloy). E) The polymer is removed leaving a metal pattern. }
\end{figure}

Samples were fabricated at the University of Maryland FabLab and the NIST Center for Nanoscale Technology using the lift-off process shown in Figure \ref{fig:A1}. Lift-off provides shape and size uniformity across large sample areas and ensures that no metal is present between the grating features.  In this study LOR-2A or PMGI (MicroChem, Newton, MA) along with S1813 (Shipley, Marlborough, MA) photoresists were used.  Substrate warping, which can complicate neutron reflectivity measurements, was minimized by selecting 1.0 mm thick substrates.  Various lithography and deposition procedures were used to optimize the fabrication process.  The procedures are detailed in Table \ref{table:AI}.
\begin{table*}
\caption{\label{table:AI}
Grating fabrication 1
}
\begin{ruledtabular}
\begin{tabular}{cccccc}
Sample & Mask Dim. & Mask Mat. & Lithography & Deposition & Resist Spacer \\
$\mu$m\_$\mu$m & $\mu$m\_$\mu$m & {} & {} & {} & {} \\
\hline
800\_800 & 800\_800 & Plastic & MJB-3 & Discovery 550 & LOR-2A \\
400\_400 & 400\_400 & Plastic & MJB-3 & Discovery 550 & LOR-2A \\
200\_200 & 200\_200 & Plastic & MJB-3 & Discovery 550 & LOR-2A \\
100\_100 & 100\_100 & Plastic & MJB-3 & Discovery 550 & LOR-2A \\
50\_50 & 50\_50 & Plastic & MJB-3 & Denton e-beam & LOR-2A \\
25\_25 & 25\_25 & Plastic & MJB-3 & Denton e-beam & LOR-2A \\
10\_10 & 10\_10 & Quartz & EVG 620 & Temescal e-beam & LOR-2A \\
5\_5 & 5\_5 & Quartz & EVG 620 & Temescal e-beam & LOR-2A \\
\end{tabular}
\end{ruledtabular}
\end{table*}

Nickel was deposited at ambient temperatures by either e-beam evaporation or radio frequency (RF) magnetron sputtering. In both techniques the sample was rotated to provide a more uniform film thickness (<2\% variation).  During e-beam deposition of the nickel film the deposition rate was 2 $\AA$/s, using a 99.995\% pure Ni Source material (Kurt J. Lesker Company).  The base pressure was $1.2\times 10^{-4}$ Pa. Sputter deposition used a Denton Discovery 550 apparatus with a 99.99\% pure nickel target (Angstrom Sciences, Duquesne, PA). The chamber base pressure was$3.1\times 10^{-4}$ Pa.  The target was pre-sputtered for 3 minutes at an argon flow rate of 50 sccm and a power of 200 Watts.  The film was deposited at an argon flow rate of 8 sccm, leading to a chamber pressure of 0.13 Pa, and a power of 50 watts providing a deposition rate of 0.07 nm/s.

Samples were characterized by stylus profilometry, optical microscopy, and scanning electron microscopy. Unpatterned thin films for each deposition instrument were characterized using x-ray reflectometry. All deposition techniques produced films with less than 3\% relative thickness variation across the wafer. All gratings were uniformly spaced and feature sizes showed minimal variation. The fabrication tools used for each process are described in Table \ref{table:AII}.
\begin{table*}
\caption{\label{table:AII}
Grating fabrication 2
}
\begin{ruledtabular}
\begin{tabular}{llll}
Equipment & Full Name & Manufacturer & Location \\
Code & {} & {} & {} \\
\hline \hline
MJB-3 & MJB-3 Mask Aligner & Karl Suss & UMD \\
EVG 620 & EVG 620 Mask Aligner & EVG & UMD \\
MA8 & MA8 Front side Contact Aligner & Suss Microtec & CNST \\
NX-2000 & NX-2000 Nano-imprint Lithography & Nanonex & CNST \\
Discovery 550 & Vacuum Discovery 550 Sputter Deposition & Denton & CNST \\
Denton e-beam & e-beam /thermal evaporator deposition & Denton & UMD \\
Temescal e-beam & e-beam deposition & Temescal & UMD \\
\end{tabular}
\end{ruledtabular}
\end{table*}

The gratings were characterized by optical diffraction measurements, primarily to establish the intrinsic correlation length of the grating stripes.  An optical reflectometer was constructed with concentric sample and detector arm rotation axes using components comparable in precision and accuracy to those  employed on the neutron reflectometer.  A laser with a nominal wavelength of 655 nm, an angular divergence of about 0.01 degrees, and a spatial width of approximately 2.3 mm in the scattering plane (at the detector aperture located 24. cm from the sample rotation axis) was used as a light source.  The linear detector aperture was placed just in front of the photocell detector with a width of approximately 0.05 mm in the scattering plane (no other apertures were used -- the exit of the laser was about 40 cm from sample center).  (Any diffraction of the light by the detector aperture, after reflection from the sample grating, was integrated over by the relatively wide photocell behind and nearly adjacent to the slit.)

\begin{figure}
\includegraphics[width=\linewidth]{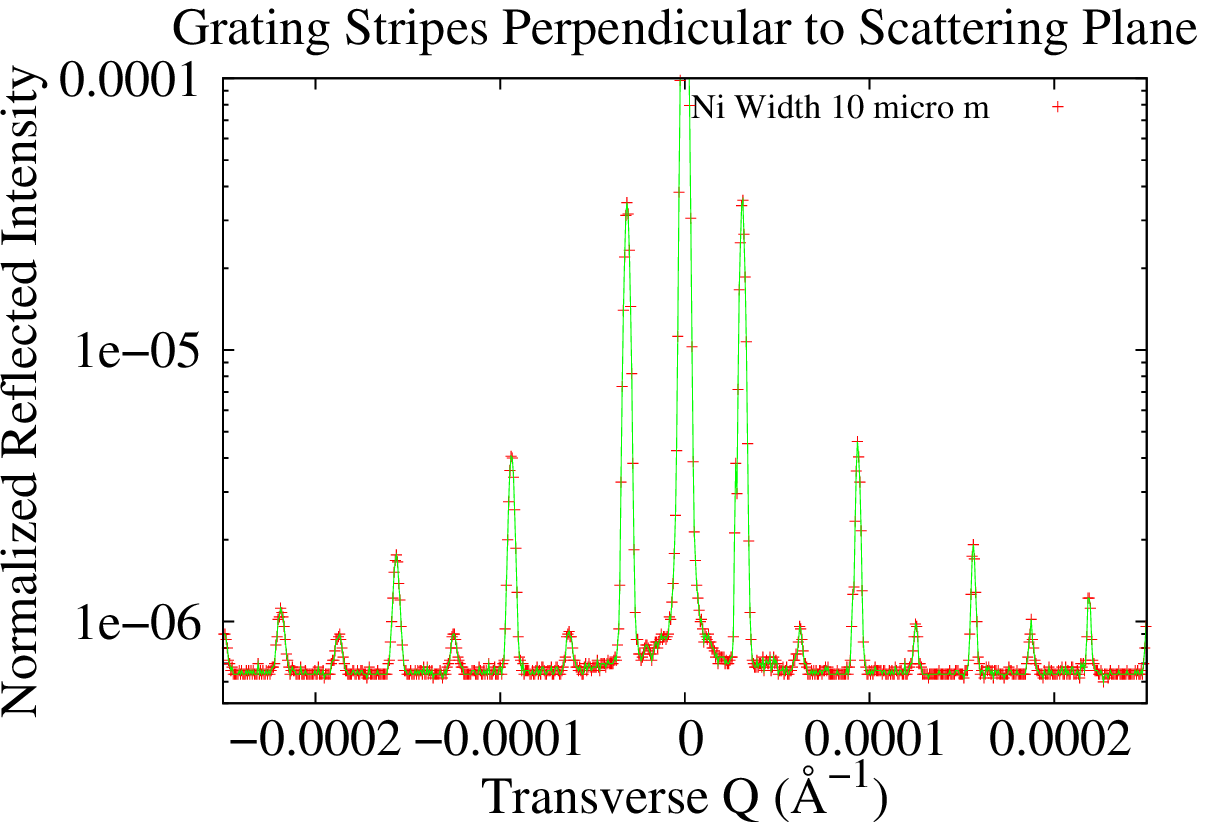}
\caption{\label{fig:A2} 
 Optical reflectometer scan for the 10 + 10 = 20 micron grating as described in the text.  The intensity of the specular peak at $Q_x = 0$ has been truncated for clarity of the nonspecular grating reflections (the even-order harmonics are suppressed because the Ni stripe width is nearly equal to the space between stripes).  The scan along $Q_x$ was done at a fixed value of $Q_z = 0.0016 \AA^{-1}$). }
\end{figure}
Without any sample, a detector (scattering angle) scan yielded an effective instrumental angular width of $\approx 0.25$ degrees.  A $Q_x$ scan at a constant value of $Q_z = 0.0016 \AA^{-1}$ was performed, the result of which is plotted in Figure \ref{fig:A2}.  
The widths of the principal reflections (m = +/- 1), higher orders, and specular ridge (m = 0) were identical, to within experimental uncertainty at a value in sample angle of approximately 0.125 degrees FWHM -- in other words, limited by the instrumental resolution.  It is of particular note that the widths of the higher order reflections did not display an increasing intrinsic width with increasing $Q$ -- which would have been indicative of a loss of correlation due to static disorder in the periodicity of the grating unit repeat structure.  The uncertainty in the line width measurement is of the order of 10 \% (upper limit).  If we conservatively attribute this amount to a natural correlation length, then deconvoluting this amount from the measured width $(3.8 \times 10^{-6} \AA^{-1})$ of the m = 1 reflection gives a minimum correlation length for the grating of about 376. microns.  This is a value significantly greater than the transverse coherence length of the neutron wave packet we are attempting to measure, even appreciably more than its corresponding projection on the surface at the low angles of reflection where the measurements were performed.  Thus the crucial assumption underlying the use of the diffraction grating structure as a measuring stick of the neutron transverse coherence length, namely that the grating period correlation length be sufficiently large in comparison, is well justified.

\clearpage
\bibliography{PRA_bibtex}

\end{document}